\documentclass{jfm}
\usepackage{graphicx}
\usepackage{epstopdf, epsfig}

\usepackage{subfig}
\usepackage{bm}% bold math
\usepackage[colorlinks=true,linkcolor=blue, citecolor=blue]{hyperref}%
\usepackage{tikz}
\usetikzlibrary{shapes}
\usetikzlibrary{positioning}
\usetikzlibrary{calc}
\newcommand{\yslant}{-0.6}
\newcommand{\xslant}{0.6}
\graphicspath{{./Figures/}}

\shorttitle{The effect of modulation on Plane Couette flow}
\shortauthor{{M. Wasy Akhtar and Rodolfo Ostilla-Mónico}}

\title{The Effect of Modulated Driving on Non-rotating and Rotating Turbulent Plane Couette Flow}

\author{M. Wasy Akhtar\aff{1},
  Rodolfo Ostilla-Mónico\aff{1,2}
  \corresp{\email{rostilla@central.uh.edu}}}

\affiliation{\aff{1}Cullen College of Engineering, University of Houston, Houston, TX 77204, USA
\aff{2}Escuela Superior de Ingener\'ia, Universidad de C\'adiz, C\'adiz, Spain}
\date{\today}

\begin{document}

%\preprint{APS/123-QED}

%\title{Modulated and Rotating Turbulent Plane Couette Flow: Direct numerical simulations}
%\author{M. Wasy Akhtar}
% \altaffiliation[Also at ]{Physics Department, XYZ University.}%Lines break automatically or can be forced with \\
%\author{Rodolfo Ostilla-Mónico}%
% \email{rostilla@central.uh.edu}
%\affiliation{Cullen College of Engineering, University of Houston, Houston, TX 77204, USA}

\maketitle

\begin{abstract}
Direct numerical simulations of turbulent non-rotating and rotating Plane Couette Flow with a periodically modulated plate velocity are conducted to study the effect of modulated forcing on turbulent shear flows. The time averaged shear Reynolds number is fixed to $Re_S = 3 \cdot 10^4$, which results in a frictional Reynolds number of approximately $Re_\tau \approx 400$. 
The modulating frequency is varied in the range $Wo\in(20,200)$, while the modulating amplitude is kept fixed at $10\%$ of the shear velocity except to demonstrate that varying this parameter changes little.
The resulting shear at the plates are found to be independent of the forcing frequency, and equal to the non-modulated baseline.
For the non-rotating simulations, two clear flow regions can be seen: a near wall region that follows Stokes' theoretical solution, and a bulk region that behaves similar to Stokes' solutions but with an increased effective viscosity. For high driving frequencies, the amplitude response follows the scaling laws for modulated turbulence of von der Heydt \emph{et al.} (Physical Review E 67, 046308 (2003)). 
Cyclonic rotation is not found to modify the system's behaviour in a substantial way, but anti-cyclonic rotation significantly changes the system's response to periodic forcing. We find that the persistent axial inhomogeneities introduced by mild anti-cyclonic rotation make it impossible to measure the propagation of the modulation adequately, while stronger anti-cyclonic rotation creates regions where the modulation travels instantaneously.
\end{abstract}


%\keywords{Suggested keywords}%Use showkeys class option if keyword
                              %display desired
\maketitle

\section{Introduction}
\label{sec:level1}

Turbulent flows subjected to periodic modulation appear at multiple scales and in many disparate contexts: pulsatile blood flow through arteries \citep{ku1997blood}, the flow of fuel, air, and other combustion products in internal combustion engines \citep{shelkin1947combustion,dent1975measurement,baumann2014validation}, and tidal currents and weather patterns in geophysical flows \citep{bouchet2012statistical,jackson1976sedimentological,turner1986turbulent}. A common feature in all such flows is that the turbulence field adjusts to the the degree of modulation, so while ordinary turbulence is often thought to have a continuum of relevant, fluctuating timescales, there is evidence that at high modulation frequencies, a dominant scale emerges that is correlated to the forcing frequency \citep{von2003response67,von2003response68,kuczaj2006response,bos2007small,kuczaj2008turbulence}. This effect can lead to phenomena such as resonances or couplings between the forcing and the existing turbulent structures that results in heavily amplified energy injection and dissipation \citep{cekli2010resonant,cekli2015stirring}.  

Experimental studies of modulated turbulence have generally been performed in wind tunnels through the use of static grids to inject energy into a flow by air streams \citep{comte1966use} or through ``active'' grids that use a grid of rods articulated by servo motors \citep{makita1991fluid}. This has allowed researchers to tune the turbulence's properties and to study the details of the dissipation rates and other features of turbulence \citep{mydlarski1996onset,poorte2002experiments}. Among other things, these studies have found that the largest energy input was reached when the time scale of the active grid forcing matched that of the largest eddies of the wind-tunnel turbulence \citep{cekli2010resonant,cekli2015stirring}. On the other hand, studies of modulated turbulence through direct numerical simulation (DNS) have been more scarce, as they require resolving all the length- and time-scales of a fully developed turbulent flow, as well as running the simulation for a sufficiently long time to capture reliable statistics. This results in high computational costs, limiting such runs to only a few studies as listed in \cite{yu2006direct} and \cite{kuczaj2006response,kuczaj2008turbulence}, which simulate randomly forced turbulence; in consonance with experiments, such studies have found resonance enhancement of mean turbulence dissipation when the forcing and flow scales match.

Compared to turbulence generated by wind-tunnels or random numerical forcing, modulated wall-bounded flows and boundary layers have received much less attention. These types of flow, which generally consist of oscillatory flows superimposed on nearly steady currents, are ubiquitous in technology and Nature. If the Reynolds number is sufficiently low, the non-linearities drop out of the Navier-Stokes equation, making the flow laminar. The solution to the problem is a linear combination of the time-dependent oscillatory solution and the steady solution, with no possibility of resonances. In this case, it may be possible to exactly solve the Navier-Stokes equations and determine the phase and amplitude of the oscillation. This provides some insights on the physics, such as the time-scales in which modulation travels through a wall-bounded flow.

The two canonical examples of laminar, modulated and wall-bounded flows are Stokes' second problem, i.e.~the flow in a semi-infinite domain driven by an oscillating wall, and pulsatile pipe flow, also known as Womersley flow \citep{womersley1955method}, i.e.~the flow in a pipe driven by an oscillatory pressure gradient. Both of these problems have exact solutions often available in textbooks \citep{landau1987fluid}, which reveal the relevant non-dimensional groups for analyzing modulated flows such as the Womersley number (which we will use below). It is also worth noting that both problems differ in the way the oscillatory component is imposed: Womersley pipe flow is driven by an oscillatory pressure gradient, which drives the bulk flow pulsations that perturb the pipe boundary layer, while in Stokes' second problem, the flow is driven by oscillating walls and hence the modulation of momentum is transported through the boundary layer towards the bulk. 

The superposition principle no longer holds in a turbulent flow. Therefore, a full simulation or experimental study is required to study the interaction of modulation with a constant flow, which may, or may not include the aforementioned resonant interactions.  
Our interest here is in flows which are driven from the boundary, which has seen less attention than modulation introduced through pressure driving \citep{scotti2001numerical,zamir2002physics,ku1983pulsatile,ling1972nonlinear}. We focus on the Plane Couette Flow (PCF) problem, the flow between two plates with a differential velocity. PCF is similar to Stokes' second problem with an additional wall that closes the system, and is ideal for studying the way a perturbation is transmitted from the wall through the boundary layer in a confined geometry. As PCF can be hard to construct experimentally, the two plates are often substituted by two cylinders, resulting in cylindrical Couette Flow, also known as Taylor-Couette flow (TCF). We note that TCF with pure inner cylinder rotation is not directly analogous to PCF, but instead to rotating Plane Couette Flow (RPCF), as the differential rotation of the cylinders is reflected as a Coriolis force unless they rotate with the same, and opposite velocities \citep{brauckmann2016momentum}. Therefore, to properly compare modulated PCF results to TCF results, solid body rotation must be added, a point to which we will return later. It is also worth noting that for certain low values of the Reynolds numbers, TCF produces a modulated response even when the driving cylinder is steady \citep{barenghi1989modulated}. We must distinguish this case, usually denoted as the modulated wavy Taylor vortex regime, from the case that interests us, i.e. fully turbulent TCF \emph{with} modulated forcing. 

Fully turbulent TCF with modulated forcing
was recently studied by \cite{VerschoofRubenA2018PdTt}, who found that the system response follows the forcing signal well for lower frequencies but falls out of phase at higher frequencies. However, they did not identify a proper time scale where the behaviour of the flow transitions from the low frequency regime to the high frequency regime. They also held the amplitude of modulation constant, and could not measure torques due to the nature of their setup. Furthermore, the effects of solid-body rotation on the response were not investigated, as the study was limited to pure inner cylinder rotation, and other rotational configurations were not considered. A proper treatment of this parameter is critical, as solid body rotation is responsible for the presence or absence of certain types of large-scale structures in the turbulent regime of rotating PCF and TCF \citep{tsukahara2010flow,salewski2015turbulent,sacco2019dynamics}, and whether they transport or not significant amounts of shear or torque through Reynolds stresses \citep{brauckmann2016momentum,kawata2019scale}. Therefore, from the previous discussions, we expect the presence or absence of large-scale structures, their physical behaviour and their interaction with the modulation to be of paramount importance in determining the response of the system when modulation is added. By modifying the rotation parameter, we can control the shape and strength of the flow structures, and provide several test cases to study the system's response to modulation including possible resonances.

While not many other studies of modulated PCF or RPCF have been conducted, modulated Rayleigh-B\'enard convection (RBC), i.e.~the flow in a fluid layer heated from below and cooled from above, has seen more attention, and can give us some hints on what behaviour to expect from modulated RPCF. This is because RBC has been shown to be in close analog to TCF \citep{busse2012twins}, with the analog to the angular momentum transport between cylinders being the heat transfer between plates, and it is another flow where large-scale dynamics heavily impact the system response.
Modulated RBC has been studied experimentally in \cite{jin2008experimental}, who found no increase in the mean heat transfer at the plates as if a sinusoidal modulation of the bottom temperature was applied. However, if the modulation was introduced through pulses or ``kicks'', a maximum heat transport enhancement of $7\%$ could be achieved when the pulse was synchronized to the existing energy scales, showing resonant enhancement in this system. \cite{jin2008experimental} rationalized this as ``spikier'' pulses being better for heat transfer enhancement than ``flatter'' ones. \cite{jin2008experimental} also found that amplitude of the fluctuations in the heat transfer and temperature were found to depend on both amplitude and frequency of the modulation in the case of pulsatile modulation. \cite{yang2020periodically} extend these results through simulations, finding a modification of the mean heat transfer of a maximum of $25\%$ in two- and three-dimensional RBC when the boundary temperature was modulated at frequencies close to the frequencies of the existing flow structures. The main difference between both cases is the amplitude of the modulation: the perturbations in the first study were  much smaller than those in the second study, which were of equal in size to the fixed temperature. 

The RBC results give us some guidelines on what we can expect when large-scale structures are present, i.e.~in the anti-cyclonic rotating regime, but there still remains a research gap on how modulated driving interacts with the general case of turbulent PCF. We will use direct numerical simulations of PCF with and without rotation to study the effect of flow modulation on wall-bounded turbulence induced by a sinusoidally oscillating wall at different frequencies and for different rotation ratios. 
We have decided to only study sinusoidal modulations to restrict the scope of this work, as the results can be directly benchmarked against the experiments of \cite{VerschoofRubenA2018PdTt}.
We will start with non-rotating PCF, and then explore cases with anti-cyclonic and cyclonic rotation to modify the large-scale structures present, and allow for different interactions of the driving with the flow. We will analyze how the flow responds to different modulation frequencies by looking at dissipation and velocity statistics. We will also study the effect of the modulating amplitude on the flow behaviour. The larger quantity and in-depth examination of available statistics will extend the findings of  \cite{VerschoofRubenA2018PdTt}, allowing us to include dissipation and spectral analysis data that was previously not available.

The paper is organized as follows: in \S 2 we describe the numerical setup (mathematical formulation, non-dimensional parameters, domain size, resolution study). In \S 3, we detail the results obtained for the non-rotating case, while in  \S 4 we add rotation and highlight the different features that arise. A brief summary and conclusions is provided in \S 5, which includes an outlook for future investigations.

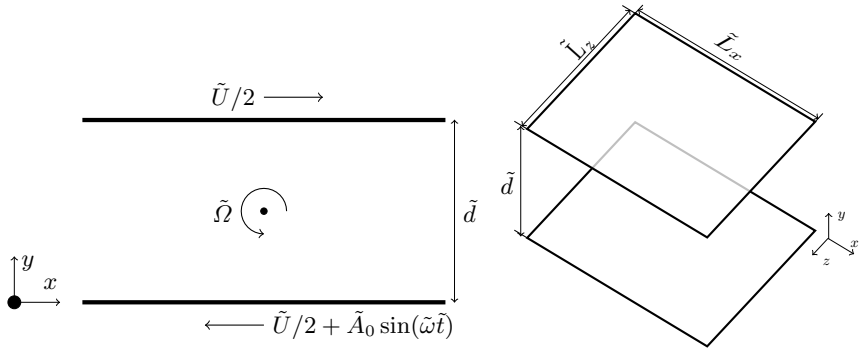
\begin{figure}
  \centering
%% Shafqat's figure
\begin{tikzpicture}[scale=0.6]
% Walls
 \draw[black,ultra thick](0,4)-- (8,4);
 \draw[black,ultra thick](0,0)-- (8,0);

% Velocity
\draw[->][black,thin](4,-0.5)--+ (-1.3,0);
\draw[->][black,thin](4,4.5)--+ (1.3,0);

 \fill[black](4,-0.5) node [scale=1,anchor=west]{$\tilde{U}/2+\tilde{A}_0\sin(\tilde{\omega} \tilde{t})$};
 \fill[black](4,4.5) node [scale=1,anchor=east]{$\tilde{U}/2$};
 
%Rotation
\draw[->][black,domain=0:270] plot ({4+0.5*cos(\x)}, {2+0.5*sin(\x)});
\filldraw[black] (4,2) circle (2pt) node[anchor=east] {};

%origin
\filldraw[black] (-1.5,0) circle (4pt) {};
%\fill[black](-1.8,0) node [scale=1,anchor= east]{$Origin$};
% Axis
 \draw[->][black,ultra thin](-1.5,0)--+ (0,1);
 \draw[->][black,ultra thin](-1.5,0)--+ (1,0);
 \draw[<->, ultra thin][black](8.2,0) to +(0,4);

% Labels:
		
 \fill[black](-0.7,0.1) node [scale=1,anchor= south]{$x$};
 \fill[black](-1.2,0.5) node [scale=1,anchor= south]{$y$};
 \fill[black](8.2,2) node [scale=1,anchor=west]{$\tilde{d}$};
 \fill[black](3.5,2) node [scale=1,anchor=east]{$\tilde{\Omega}$};
\end{tikzpicture}
\begin{tikzpicture}[scale=0.34]
% Bottom Wall
\begin{scope}[ yshift=-120, every node/.append style={yslant=\yslant,xslant=\xslant},
		yslant=\yslant,xslant=\xslant
	] 
		% The frame:
		\draw[black, thick] (0,0) rectangle (7,7);
		
		% Extra line for dimensions
        \draw[black](0,0) to (-0.5,0);
        
        % Axis
        \draw[->, ultra thin][black](7.5,7) to +(1,0);
        \draw[->, ultra thin][black](7.5,7) to +(0,-1); 
        
	\end{scope}
	
	% vertical lines for dimension wall gap
	
    \draw[<->][black,ultra thin](-0.25,0.1,0) to (-0.25,-4.15,0);
    
    \draw[->, ultra thin](7.49,-8.5,-11) to + (0,1,0);

	%Labels
	\fill[black](xyz cs:x=-0.2,y=-2.1,z=0) node [scale=1,anchor= east]{$\tilde{d}$};
	\fill[black](8,-8,-11) node [scale=0.6,anchor= south]{$y$};
	\fill[black](8.5,-9,-11) node [scale=0.6,anchor= south]{$x$};
	\fill[black](8.1,-9,-9.2) node [scale=0.6,anchor= south]{$z$};
	
% Top Wall
	\begin{scope}[
		yshift=0,
		every node/.append style={yslant=\yslant,xslant=\xslant},
		yslant=\yslant,xslant=\xslant
	]
		% The frame:
		\fill[white,fill opacity=.75] (0,0) rectangle (7,7); %
		\draw[<->,ultra thin][black](-0.2,0) to (-0.2,7);
		\draw[black](0,0) to (-0.5,0);
%		\draw[black](7,0) to + (0.5,0);
		\draw[black](0,7) to + (-0.5,0);
		
		\draw[<->,ultra thin][black](0,7.2) to (7,7.2);
		\draw[black](0,7) to + (0,0.5);
		\draw[black](7,7) to + (0,0.5);
		Opacity
		\draw[black, thick] (0,0) rectangle (7,7); 

		 % Labels:
		
		\fill[black](-0.7,3.5) node [rotate=90, scale=1,anchor= west]{$\tilde{L}_z$};
		\fill[black](3.2,7.2) node [ scale=1,anchor= south]{$\tilde{L}_x$};
		
	\end{scope} 
\end{tikzpicture}

  \caption{Schematic of the (dimensional) simulation domain. The third (spanwise) dimension $z$ is omitted in the left panel for clarity.} 
  \label{fig:Taylor_simulation_setup}
\end{figure}

%% Waleffe section

\section{Numerical setup}

To simulate RPCF we use the three-dimensional Cartesian domain shown in Fig. \ref{fig:Taylor_simulation_setup}. The top and bottom plates have length $\tilde{L}_{x}$ (streamwise) and $\tilde{L}_{z}$ (spanwise) and are separated by gap width $\tilde{d}$. Periodic conditions are imposed at the $x$ and $z$ domain boundaries. Solid body rotation in the $z$-direction is added through a Coriolis force. This represents the differential motion of the cylinders in a Taylor-Couette system as the curvature vanishes \citep{brauckmann2016momentum}. 

The top and bottom plates are no-slip, and prescribed to have opposite streamwise velocities $\pm (\tilde{U}/2)\textbf{e}_x$, where $\textbf{e}_i$ is the unit vector in the $i$-direction. In addition to this steady shear, a modulation is superimposed onto the bottom plate's velocity with a perturbation frequency $\tilde{\omega} = 2 \pi / \tilde{T}$ and magnitude $\tilde{A}_0$ such that the total velocity at the bottom plate is $-\tilde{U}/2+\tilde{A}_0\sin(\tilde{\omega}\tilde{t})$, with $\tilde{t}$ and $\tilde{T}$ the dimensional time and period respectively.

The incompressible Navier-Stokes equations are made dimensionless using the gap-width $\tilde{d}$ and the plate velocity $\tilde{U}$. They read:

\begin{equation}
 \frac{\partial \textbf{u}}{\partial t} + \textbf{u}\cdot \nabla \textbf{u}  + R_\Omega (\textbf{e}_z \times \textbf{u}) = -\nabla p + Re_s^{-1} \nabla^2 \textbf{u},
  \label{eq:ns1}
\end{equation}
 
\noindent which alongside the incompressibility condition defines the flow field,
 
\begin{equation}
 \nabla \cdot \textbf{u} = 0.
 \label{eq:ns2}
\end{equation}

\noindent Here $\textbf{u}$ is the non-dimensional velocity, $p$ the non-dimensional pressure and $t$ is the dimensionless time $t=\tilde{t}U/d$. Equation \ref{eq:ns1} contains two non-dimensional control parameters: a shear Reynolds number, $Re_s=\tilde{U}\tilde{d}/\nu$ and the Coriolis parameter (sometimes known as the Rotation number), $R_{\Omega}=2 \tilde{\Omega}\tilde{d}/\tilde{U}$ with $\tilde{\Omega}$ the background spanwise rotation,. Two more control parameters are provided by the modulated boundary condition: the non-dimensional modulation amplitude $\alpha=\tilde{A}_0/\tilde{U}$ and the non-dimensionalized modulation frequency, which is written as the Womersley number ($Wo$) defined as $\tilde{d} \sqrt{\tilde{\omega}/\nu}$ following \cite{VerschoofRubenA2018PdTt}. Using this, the dimensionless streamwise velocity boundary conditions become $\textbf{u}=\frac{1}{2}\textbf{e}_x$ at the top plate and 

\begin{equation}
 \textbf{u}=\frac{1}{2}+\alpha\sin\left (\frac{t}{T} \right ) \textbf{e}_x=\frac{1}{2}+\alpha\sin\left (\frac{Wo^2}{2\pi Re_s}t \right) \textbf{e}_x
\end{equation}

\noindent at the bottom one, with $T=2\pi Re_s/Wo^2$ the non-dimensional period. 

Periodic aspect ratios of $L_x=\tilde{L}_x/\tilde{d}=2\pi$ and $L_z=\tilde{L}_z/\tilde{d}=\pi$ are used. The Reynolds number $Re_s$ is fixed at $3 \times 10^4$, resulting in a frictional Reynolds number $Re_\tau= u_\tau \tilde{U} \tilde{d}/(2\nu)= u_\tau Re_s/2 \approx 400$ for the non-rotating case, where $u_\tau$ is the non-dimensional shear velocity defined as $u_\tau = Re_s^{-1/2} \sqrt{\partial_y \langle u_x(y=0) \rangle_{A}}$, where $\langle . \rangle_{A}$ denotes averaging with respect to time and to the streamwise and spanwise directions. For convenience, we also define a dimensionless frictional time unit $t_\tau=\tilde{t} u_\tau \tilde{U}/\tilde{d} = t u_\tau$ which will become useful later. We note that $t_\tau$ is a diagnostic time which does not appear in our equations, as $u_\tau$ is dynamically determined. 

The Rotation number $R_{\Omega}$ is varied in the range $[-0.1, 0.3]$, with positive values denoting anti-cyclonic rotation, such that the spanwise rotation vector is anti-parallel to the vorticity of base flow, whereas negative values of the Coriolis signify cyclonic behavior, i.e. the spanwise rotation is parallel to the vorticity vector of the base flow. The perturbation amplitude $\alpha$ is kept constant at $\alpha=0.1$ unless stated otherwise. The Wormersley number $Wo$ is varied in the range $Wo\in [26,200]$, with selected cases at higher $Wo$. Table \ref{tbl:wos} shows how these values of $Wo$ correspond to the different time-scales in the flow, including the dimensionless forcing period in both dimensionless time units $t$ and $t_\tau$.

\begin{table}
    \centering
\begin{tabular}{|c|c|c|}
\hline
% $Wo$ & $T$ & $T_\tau$ & $2T_\tau$ \\
% \hline
% 26 & $2.66\times10^2$ & $7.15\times10^0$ & $1.43\times10^{1}$ \\
% 44 & $9.61\times10^1$ & $2.49\times10^0$ & $5.00\times10^{0}$ \\
% 77 & $3.18\times10^1$ & $8.15\times10^{-1}$ & $1.63\times10^{0}$ \\ 
% 114 & $1.44\times10^1$ & $3.72\times10^{-1}$ & $7.44\times10^{-1}$ \\
% 200 & $4.71\times10^0$ & $1.20\times10^{-1}$ & $2.41\times10^{-1}$ \\
% 300 & $2.09\times10^0$ & $5.37\times10^{-2}$ & $1.07\times10^{-1}$ \\
% 400 & $1.18\times10^0$ & $3.02\times10^{-2}$ & $6.04\times10^{-2}$ \\
$Wo$ & $T$ & $T_\tau$ \\
\hline
26 & $2.66\times10^2$ & $1.43\times10^{1}$ \\
44 & $9.61\times10^1$ & $5.00\times10^{0}$ \\
77 & $3.18\times10^1$ & $1.63\times10^{0}$ \\ 
114 & $1.44\times10^1$ & $7.44\times10^{-1}$ \\
200 & $4.71\times10^0$ & $2.41\times10^{-1}$ \\
300 & $2.09\times10^0$ & $1.07\times10^{-1}$ \\
400 & $1.18\times10^0$ & $6.04\times10^{-2}$ \\
\hline 
\end{tabular}
    \caption{Summary of Wormersley numbers used for all simulations, and their corresponding dimensionless forcing period $T=\tilde{T}U/d$. We also include the forcing period in dimensionless frictional units $T_\tau$ defined as $T_\tau=(\tilde{T} u_\tau U)/(d/2)=2Tu_\tau$ to show how $T$ relates to the frictional time-scales in the flow.}
    \label{tbl:wos}
\end{table}

The equations Eq.~\ref{eq:ns1}-\ref{eq:ns2} are discretized using finite differences: second-order accurate energy conserving in space, third-order accurate in time using Runge-Kutta for the explicit terms. The viscous term is discretized in the wall normal direction using a second-order Crank-Nicholson scheme. The discretized equations are solved using the parallel FORTRAN based code, AFiD (\url{www.afid.eu}). This code has been used in previous studies to study turbulent Rayleigh-B\'enard convection and Taylor-Couette flow \citep{van2015pencil} and has been thoroughly validated. Details of the code algorithms are documented in \cite{verzicco1996finite} and \cite{van2015pencil}.  Spatial resolution of the simulations are selected as $N_x \times N_y \times N_z = 512\times 384\times512$ in the streamwise, wall-normal and spanwise directions, respectively. The points are distributed uniformly in the streamwise and spanwise direction, while for the wall-normal direction they are clustered near the walls using a clipped Chebychev distribution. This gives us an effective resolution in viscous wall units of $\Delta x^+=9.8$, $\Delta z^+=4.9$ and $\Delta y^+ \in(0.3,3.0)$. This resolution is chosen in accordance with the spatial resolution selected for $Re_s=3.61 \times 10^4$ in the study of turbulent Taylor rolls in \cite{sacco2019dynamics}. For a series of selected cases, we double $L_x=4\pi$ and $L_z=2\pi$, to check the dependence of the statistics on the box-size. For these cases, we also double the resolutions in the $x$ and $z$ directions to keep the same base grid spacing.

A variable time-stepping scheme is defined such that the maximum CFL condition does not exceed 1.2. To exclude the start-up transients, the first two hundred time units are discarded before starting to evaluate the statistics. The duration of the simulation to evaluate turbulence statistics is 10 periods from the end of the simulation (except for $Wo=26$, for which it is 5 due to long run times) for a given $Wo$ number, or one thousand simulation time units, which ever is larger. 

Temporal convergence is also checked by monitoring that the $y$-dependence of the computed non-dimensional momentum flux

\begin{equation}
    J^\omega = \langle u_x u_y \rangle_{A} -  Re_s^{-1} \displaystyle\frac{\partial \langle u_x \rangle_{A}}{\partial y}
\end{equation}

\noindent does not exceed 1\%. While $J^\omega$ is independent of $y$ for sufficiently long averaging times, in our simulations it will show a $y$ dependence due to finite averaging times. Therefore, quantifying the $y$-dependence of $J^\omega$ is a way to assess the errors made due to finite statistics. We note that performing this check includes checking that the shear at both walls is equal to within $1\%$, as the shear at the walls is simply $J^\omega$. 

\section{Results for non-rotating Plane Couette Flow}
\label{sec:dissro0}

The first case we analyze is Plane Couette flow without rotation, i.e.~$R_\Omega=0$. Non-rotating Plane Couette flow contains large-scale structures that extend significantly in the streamwise direction \citep{tsukahara2006dns}. However, these structures do not dominate the transport of momentum in the same way that the Taylor rolls present for $R_\Omega=0.1$ \citep{brauckmann2016momentum,sacco2019dynamics,sacco2020dynamic}. Instead, momentum is transferred through a hierarchy of eddies which spans many length- and time-scales \citep{townsend1980structure}, so in principle we do not expect that there are natural time-scales in the flow with which the modulation could couple to produce resonances.

We first look at the volumetrically averaged instantaneous dissipation $\varepsilon$, which is a quantity of interest in studies of modulated turbulence. We note that the temporal average of $\varepsilon$ is equal to the shear force at the plates (modulo scaling factors) due to the exact balances of energy: in the statistically stationary state on the average energy input through the walls must be balanced out by the viscous dissipation. 

To represent this, we follow \cite{brauckmann2016momentum} and \cite{eckhardt2020exact} to define two Nusselt numbers. First, we define a force Nusselt number $Nu$ as $Nu=J^\omega/J^\omega_{lam}$, where $J^\omega_{lam}$ is the momentum current for the laminar state. By definition $Nu=1$ in the purely streamwise flow. Second, we define a Nusselt number $Nu_\varepsilon$ based on the instantaneous viscous dissipation, $Nu_\varepsilon = \varepsilon/\varepsilon_{lam}$. We note that due to the exact balances, while the instantaneous value of $Nu_\varepsilon$ will be distinct from instantaneous force Nusselt number $Nu$, the time-averaged value of $Nu_\varepsilon$ will equal to $Nu$.

\begin{figure}
  \includegraphics[width=.32\textwidth]{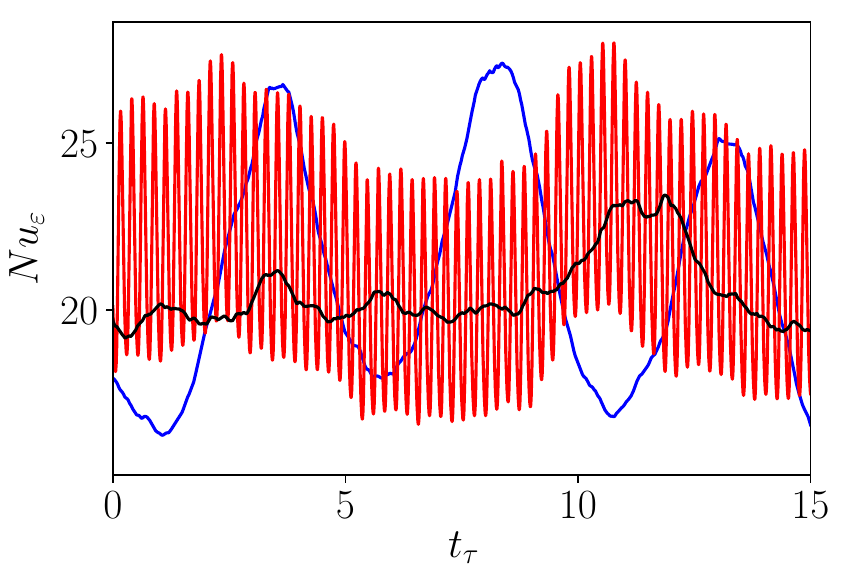}
  \includegraphics[width=.32\textwidth]{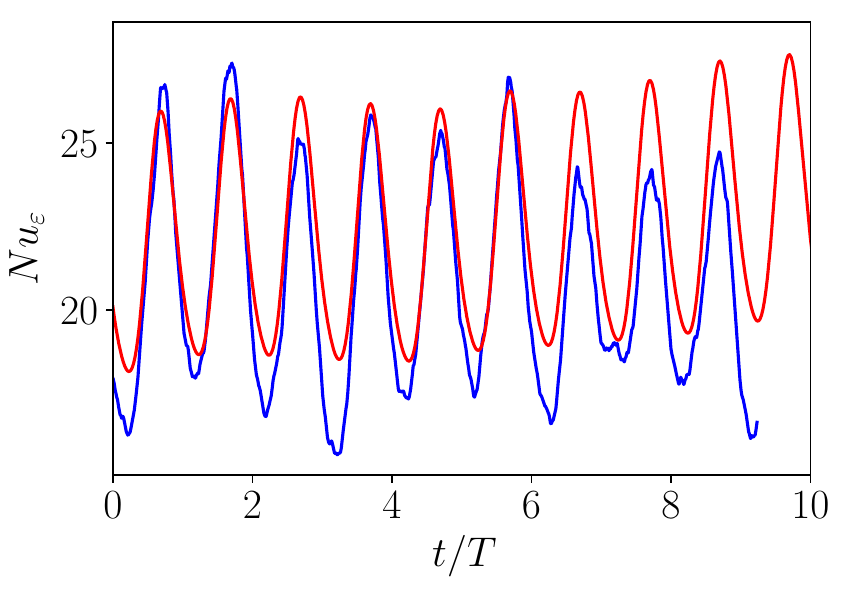}
  \includegraphics[width=.32\textwidth]{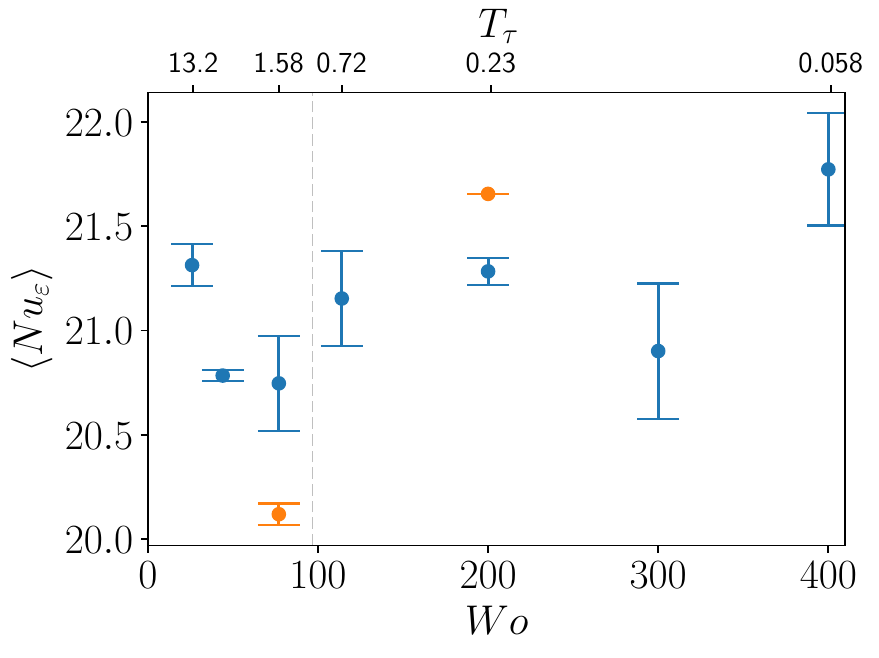}
\caption{Left panel: Temporal evolution of the averaged dissipation, non-dimensionalized as a Nusselt number ($Nu_\varepsilon$) for unmodulated flow (black), $Wo=44$ (blue) and $Wo=200$ (red). Center panel: Same as the left panel, with the time units re-scaled using the period of the forcing. Right panel: Temporally averaged $Nu$ against $Wo$ for the non-rotating case. Blue points denote the baseline periodic aspect ratios of $L_x=2\pi$ and $L_z=\pi$, while the orange data points are simulations to check the effect of domain size with $L_x=4\pi$ and $L_z=2\pi$.}
\label{fig:diss}
\end{figure}

In the left panel of Figure \ref{fig:diss}, we show the instantaneous values of $Nu$ for two different values of $Wo$ (44, 200) and for the unmodulated flow. In the left panel, time is non-dimensionalized using the frictional time-scale. In the center panel we use the modulation period $T$ to non-dimensionalize time, and show only the instantaneous values of $Nu$ for the modulated cases. This choice rescales the horizontal axis differently for both lines, and reveals how there are two clear time-scales in the flow, one given by the modulated forcing from the wall, and of the order $\mathcal{O}(T)$, and one given by the turbulent flow itself, of the order of $\mathcal{O}(t_\tau)$. The fluctuations due to natural turbulence appear to be much smaller than those introduced by modulation (left panel), which are of the order of $20-30\%$ around the mean value of $Nu$ even when the wall-modulation amplitude is only 10\% of the average wall velocity. We also note that the fluctuation size does not appear to change appreciably with the modulation frequency, unlike what was observed in \cite{jin2008experimental}, a fact which we will return to later. 

%% Nusselt vs t/T and Wo

To elucidate how the average value of dissipation (and wall-shear) depends on the parameters of the unsteady forcing, we show the temporally averaged values of $Nu$ in the third panel of Figure \ref{fig:diss}. We cannot observe any definite patterns in the resulting values for total dissipation: they deviate from the value of $\langle Nu\rangle =20.6\pm0.2$ obtained with no modulation, but this baseline value is generally contained within the error bars of the simulation. This provides a first indication that the modulation does not significantly couple with any existing structures in the flow, even for $Wo=77$ and $Wo=114$, when the modulation roughly matches the time-scale of the flow, i.e. $T_\tau\approx 1$. To assess the possible effects of box-size dependence on these results, we simulated two additional cases at twice the domain size for $Wo=77$ and $Wo=200$, shown with orange markers in the panel. These values of $Nu$ also show some dispersion around the baseline value. We note that the $Wo=77$ case is $4\%$ below the previous value, which would result in a smaller friction at the walls and is consistent with similar studies of PCF \citep{tsukahara2006dns}. However, this is not seen for $Wo=200$, where the resulting $Nu$ is instead larger. We conclude by stating that the error introduced by the small domain size is comparable to or larger than any variation due to $Wo$, which means we cannot make any definite statements on the $Nu(Wo)$ dependence; even if following  \cite{jin2008experimental}, we do not expect there to be an effect.

\begin{figure}
  \includegraphics[width=.48\textwidth]{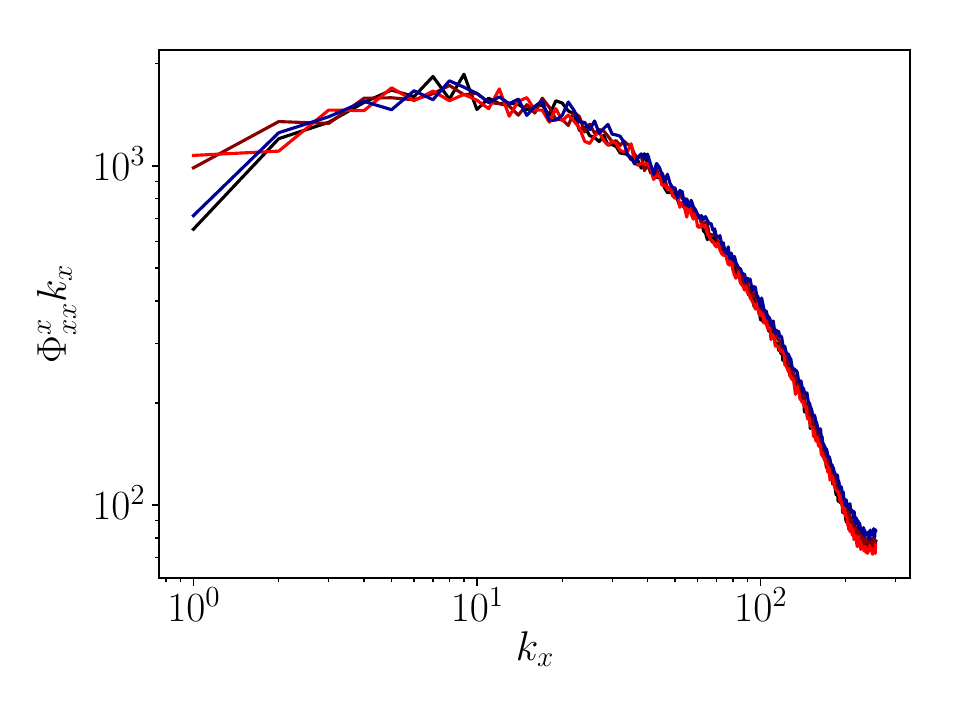}
  \includegraphics[width=.48\textwidth]{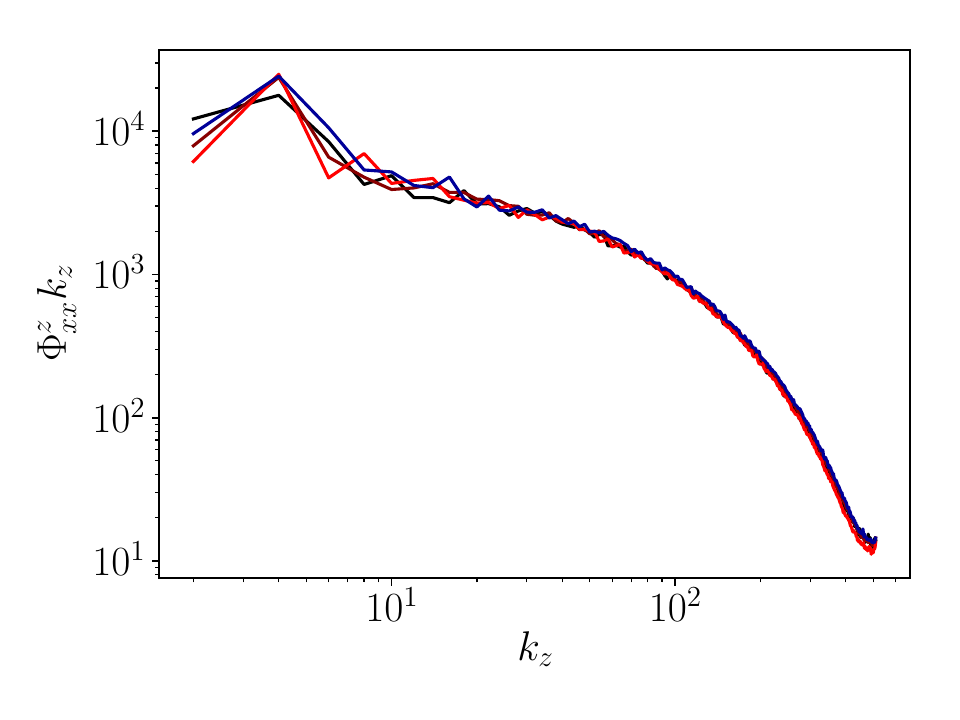}
\caption{Time-averaged pre-multiplied streamwise (left) and spanwise (right) spectra of the streamwise velocity at the mid-gap ($y=0.5$) for non-rotating PCF. Symbols: $Wo=44$ (black), $Wo=77$ (dark red), $Wo=200$ (light red), unmodulated (dark blue).}
\label{fig:specro0}
\end{figure}

We first check the effect of modulation on the flow structures by showing the streamwise and spanwise spectra of the streamwise velocity in Figure \ref{fig:specro0}. The periodic modulation does not introduce significant modifications of the energy spectra at the mid-gap. Only a small degree of variation between the cases, especially at the low wavenumber end can be seen, and among these there is no discernible pattern of behavior as the curves are not ordered by $Wo$. This is similar as to what was seen for $Nu$, where no discernible pattern could be seen as $Wo$ was changed. We also note that for the streamwise spectra $\Phi_{xx}^x$, the unmodulated case is closest of all to the lowest $Wo$ curve ($Wo=44$), something which is unexpected. Due to the absence of obvious patterns, we may attribute these differences to insufficient temporal convergence of the statistics shown in the graphs.

%% Nusselt vs t/T and Wo

\begin{figure}
  \includegraphics[width=.45\linewidth]{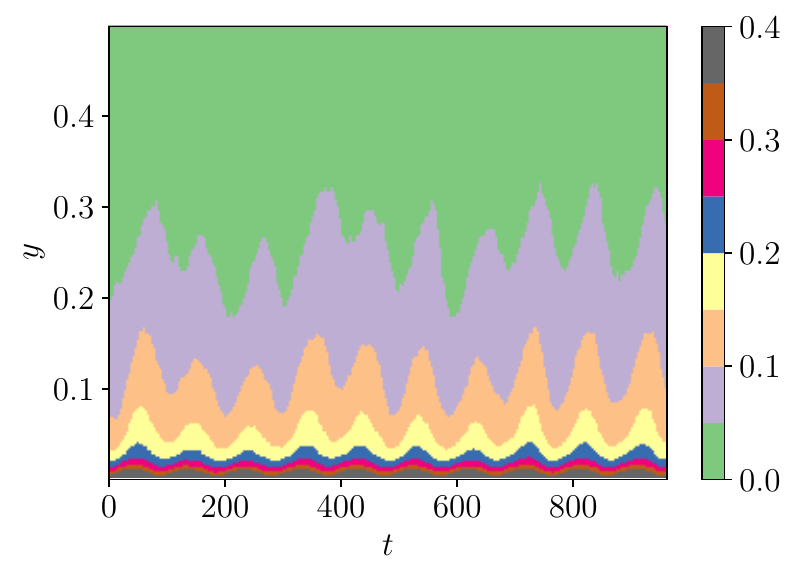}
  \includegraphics[width=.45\linewidth]{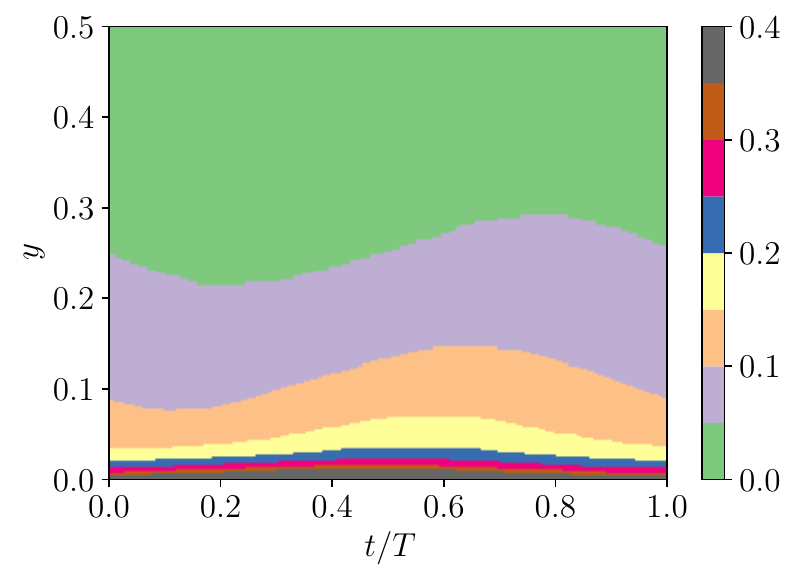}
\caption{Space-time pseudocolor plots of the averaged streamwise velocity up to the mid-gap for $Wo=44$, non-rotating case. Left: Instantaneous stream- and span-wise averaged  velocity. Right: Spanwise, streamwise and phase averaged velocity. }
\label{fi:averagev}
\end{figure}

To understand how the modulation is transferred through the flow, we turn towards the streamwise velocity field itself. To isolate the effect of the modulation from the turbulent background fluctuations, we first average the field in the span- and stream-wise directions. A space-time visualization of the result in shown in the left panel of Figure \ref{fi:averagev}. The modulation imposed by the unsteady boundary condition can be clearly seen. Due to the turbulent fluctuations, the average velocities are not periodic. To separate the effect introduced by the periodic motion of the wall, we conduct a phase average over several of the periods simulated. This reduces the temporal domain to $0\leq t/T < 1$, resulting in a phase-averaged velocity field that we denote as $\bar{u}(y,t/T)$. We show a space-time visualization of $\bar{u}$ in the right panel of Figure \ref{fi:averagev}, where we can now clearly see how the modulation wave travels from the wall into the rest of the fluid, observing that as the distance from the wall increases, the phase lag becomes larger. 

%Fig 3
%% Phase delay vs. y for Ro = 0

We can decompose $\bar{u}$ in the following manner:
\begin{equation}
 \bar{u}(y,t/T) = \bar{u}_0(y) + f(y,t/T),
 \label{eq:fdefin}
\end{equation}

\noindent where $\bar{u}_0$ is the temporally averaged velocity, and $f(y,t/T)$ denotes the periodic effect introduced by the modulation. 
This manuscript will focus on the behaviour of  $f(y,t/T)$, but before we do so, we wish to mention that the behaviour of $\bar{u}_0$ is not very different from that seen in unmodulated rotating Plane Couette flow. While the basic symmetry is broken, it is restored on a temporally-averaged sense, and as a consequence, when we examine the behaviour of the average streamwise velocity for example, we cannot observe an asymmetry between the modulated and unmodulated wall.

Returning to $f$, there is no a priori reason to think that it cannot contain any harmonics of the fundamental modulation of period $T$, i.e. $T/2$, $T/3$, etc. However, the visualization shown in the right panel of Figure \ref{fi:averagev} indicates that the dominant temporal scale is that associated with the modulation, and not to higher harmonics. This is further confirmed by Fourier analysis, which shows that the second harmonic has an amplitude that is a factor $10-300$ times smaller than the first fundamental, depending on the distance to the wall. 

\begin{figure}
\includegraphics[width=0.49\textwidth]{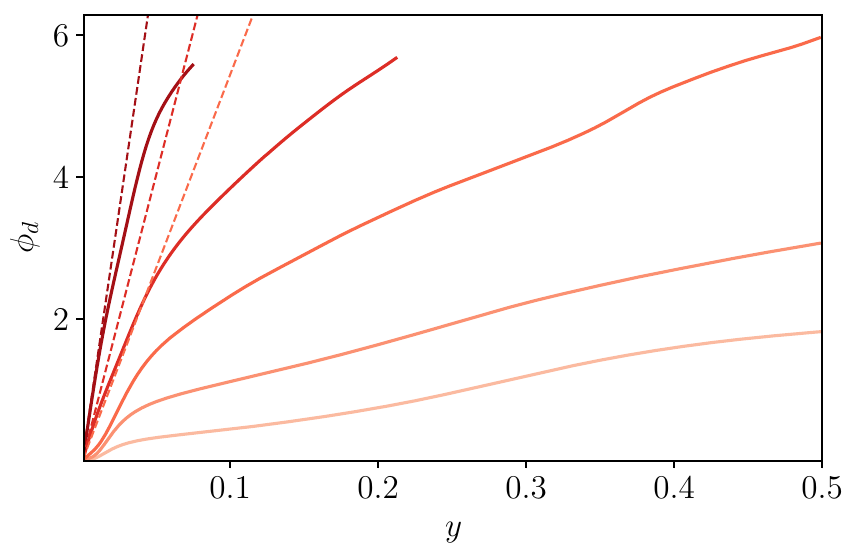}
\includegraphics[width=0.49\textwidth]{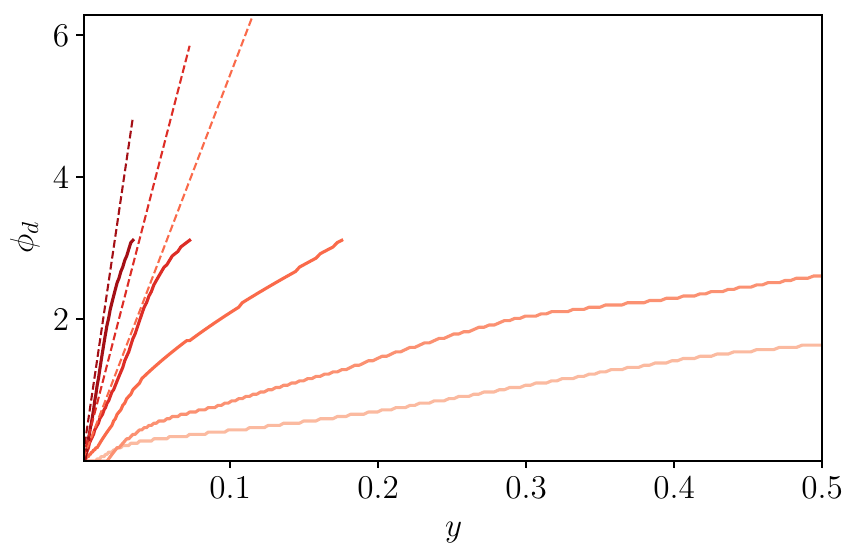} \includegraphics[width=0.49\textwidth]{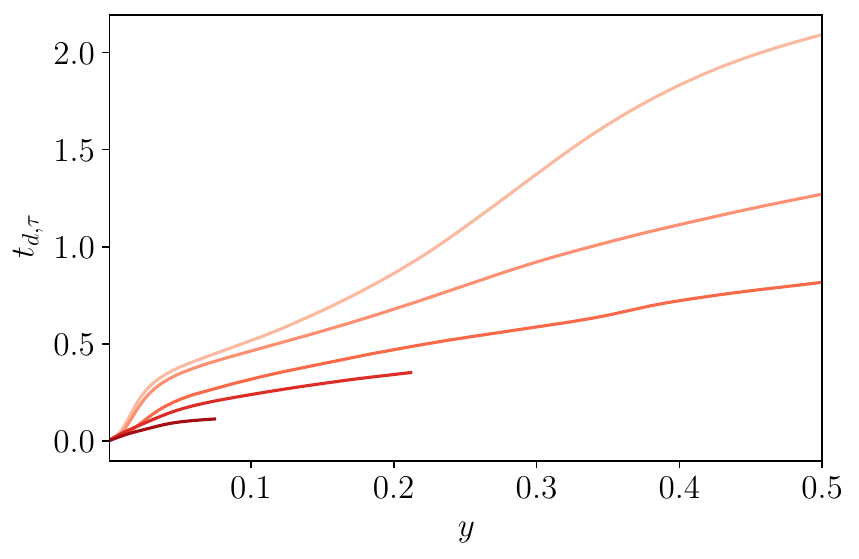}
\includegraphics[width=0.49\textwidth]{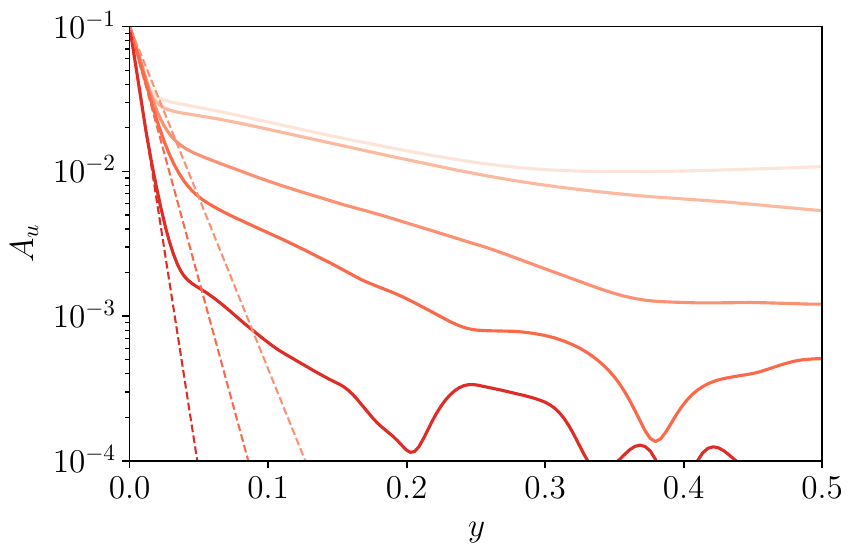}
\caption{Top panels: Phase delay $\phi_{d}$ against wall distance for the non-rotating cases measured through FFT (left) and using the cross-correlation method of \cite{VerschoofRubenA2018PdTt} (right). Bottom left: Perturbation time lag with respect to wall against wall distance for non-rotating cases. Bottom right: Amplitude response $A_{u}$ measured through the FFT method against wall distance for the non-rotating cases. Symbols: $Wo=26$, $44$, $77$, $114$, $200$ from light red to dark red. Dashed lines in select panels are the theoretical results from Stokes' problem. }
\label{fig:phaselag}
\end{figure}

Therefore, we use the ansatz that $f$ has the following functional dependence:

\begin{equation}
    f(y,t/T) = A_u(y) \sin[2\pi t/T + \phi_{d}(y)],
\end{equation}

\noindent where $A_u$ is the amplitude response, and $\phi_{d}$ the phase lag, both of which are dependent on the distance to the wall. To determine the values of these quantities we use two methods. The first is to simply take a Fourier transform of $f$, and determine $\phi_{d}$ and $A_u$ from the results of this transform. We truncate $\phi_d$ when the amplitude of the Fourier mode is smaller than $10^{-3}$ for reasons that will become apparent below. For comparison purposes, we also follow \cite{VerschoofRubenA2018PdTt}, and determine the phase-delay using the peak of the cross-correlation between the wall-velocity and $f$. We show the results obtained from both methods in the top panels of Figure \ref{fig:phaselag} for values of $Wo$ in the (26,200) range. We have also added dashed lines which represent the exact solution for Stokes' oscillating laminar boundary layer, $\phi_{d}=(y~Wo)/\sqrt{2}$.

%Fig 5
%% Phase delay vs. y for Ro = 0

We first notice that the phase delay results for the cross-correlation are limited to the range $[0,\pi)$, while those obtained from the Fourier transform have a larger range of $\phi$ which is only limited by truncation. The results are also qualitatively similar to each other, with a few minor differences. In the cross-correlation method, the phase delay is effectively zero for the lowest $Wo$ very close to the wall, and is slightly smaller than the one obtained through the Fourier transform. This method also shows some ``graininess'' due to the numerical inadequacies of using the maximum operator. Hence, we use the Fourier transform for measuring $A_u$ and $\phi_{d}$, as the trend is more distinct, even if it is evident that the cross-correlation calculation used in ~\cite{VerschoofRubenA2018PdTt} also gives reasonable results.

Turning to the results themselves, we can observe there are two distinct regions where the phase and the wall distance are related in an approximately linear manner, albeit with different slopes. A linear relationship between phase and wall distance can be understood as information from the modulation propagating into the flow with a constant speed, with the phase delay acting as a proxy for time ($\phi_d\sim t/T$). In this way, a steeper slope $m$ in the $\phi_d=my$ line signals a slower travel velocity. The lines change from one slope to another at $y\approx 0.05$, which corresponds to $y^+\approx 40$, i.e. the buffer sub-region of the boundary layer.
Hence, the region $y \le 0.05$ approximately corresponds to the viscid sub-region where the modulation travels slower, mainly through viscosity. The other flow region ($y \ge 0.05$) corresponds to zones in the buffer layer and beyond. It has a shallower slope, which means that the modulation travels faster, as it is essentially transported by turbulent fluctuations.

In the viscous region ($y\le 0.05$), the slopes obtained are shallower than the purely viscid solutions (denoted as dashed lines), so other mechanisms that accelerate the transport are at play. We also highlight that the distance between the purely viscous solution and the actual solution decreases as the frequency increases, meaning that these corrections become less important.
We can also observe that the phase delay is larger with increasing $Wo$ at a given wall distance. This does not mean that the perturbation itself travels slower. Instead, this means that increasing the frequency of the perturbation does not increase the travel speed of the perturbation in a sufficient amount to make the phase delay constant at a given distance. To emphasize this point, we show the actual delay time in frictional time-units $t_{d,\tau}$ against wall distance for all values of $Wo$ in the bottom left panel of Figure \ref{fig:phaselag}. Again, we can see two regions with different behavior: the viscous subregion at $y\le  0.05$ and the turbulent region for $y>0.05$. In the near wall region, if $Wo$ is large (dark curves), the perturbation travel speed is fast and strongly depends on $Wo$. If $Wo$ is small (light curves), the travel speed of the perturbation is slower and the two curves seem to almost lie on top of each other for small values of $y<0.03$. This is consonant with the fact that high-$Wo$ curves follow the Stokes' solution better than their low-$Wo$ counterparts in the viscid region.

We can observe similar behaviour in the turbulent region, where all lines reported (except for $Wo=26$) have a similar slope indicating that the actual velocity at which the perturbation travels is approximately $Wo$-independent in the bulk. The transition between high-$Wo$  and low-$Wo$ behavior is not easy to delimit, as some curves show different characteristic behavior depending on the wall distance. For $Wo=26$, due to low travel speed, we do not expect a simple picture. Once the delay time grows beyond $t_\tau=1$, the information from more than one modulated cycle will be affecting the flow. Finally, we notice that for $Wo=77$, the phase delay is approximately $2\pi$ at the center; in contrast for $Wo=44$, it takes approximately $t_\tau=1$ time units for the perturbation to reach the center. This confirms the fact that we are forcing close to the natural time-scale of the flow as suggested by Table \ref{tbl:wos}. However, we do not see any sort of resonant behaviour in the dissipation--neither for $Wo=77$ nor for $Wo=44$.

We now turn to the amplitude response $A_u$. We only show results here obtained through the Fourier transform method, which presents much smaller oscillations than applying the method used in \cite{VerschoofRubenA2018PdTt} on our data. In the bottom right panel of Fig.~\ref{fig:phaselag}, we show $A_u$ as a function of wall-distance for the same five $Wo$. We also include the solutions of Stokes' second problem for the three largest values of $Wo$, which is given by $A_u=\exp(-\sqrt{2}~Wo~y)$. These are represented as straight lines in our semi-logarithmic plot. The same two regions as in the phase plot can be seen: There is an inner viscid region where the perturbation amplitude decays rapidly, and which closely tracks the viscous solution. There is also a turbulent region where the perturbation decay is slower. And again, the  transition between both regions happens at $y\approx0.05$.  When $y>0.05$, the perturbations are transported through turbulence, which can be understood as an effective increase of the viscosity which in turn facilitates the propagation of the perturbation effectively resulting in a slower decay rate. For the case of $Wo=26$, we can observe a third region that starts at around $y=0.3$ where the slope further decreases to the point that the amplitude is almost constant. The origin of this third region is unclear, as similar changes did not clearly appear in the phase delay but were present when looking at the delay time. In principle, we can rule out averaging errors due to the rather large magnitude of $A_u$. Finite-averaging and other numerical errors are expressed in the manner seen for the two highest values of $Wo$: through oscillations which only start to dominate once $A_u<10^{-3}$, i.e.~as the perturbation amplitude has decreased by $\mathcal{O}(10^2)$. A possible source of this could be that in this region the delay time from the wall exceeds a modulation cycle, causing more complicated interactions. 

\begin{figure}
\includegraphics[width=.45\linewidth]{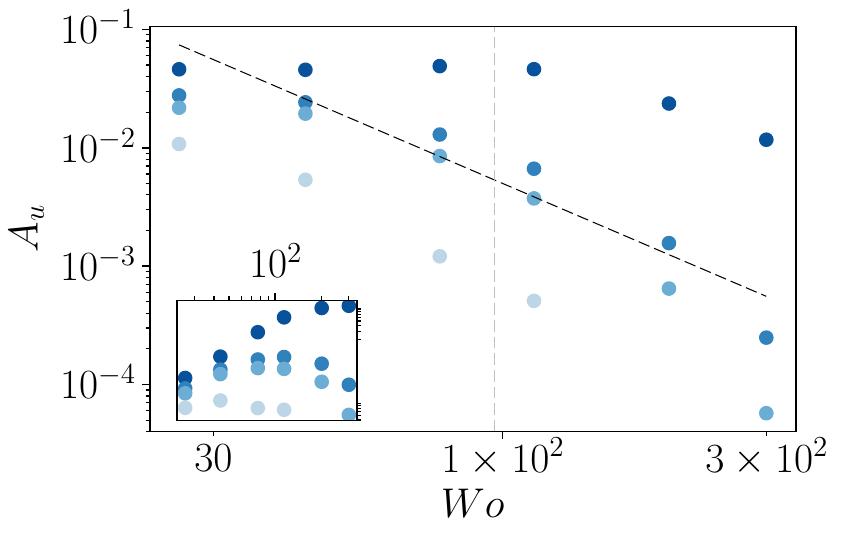}
\includegraphics[width=.45\linewidth]{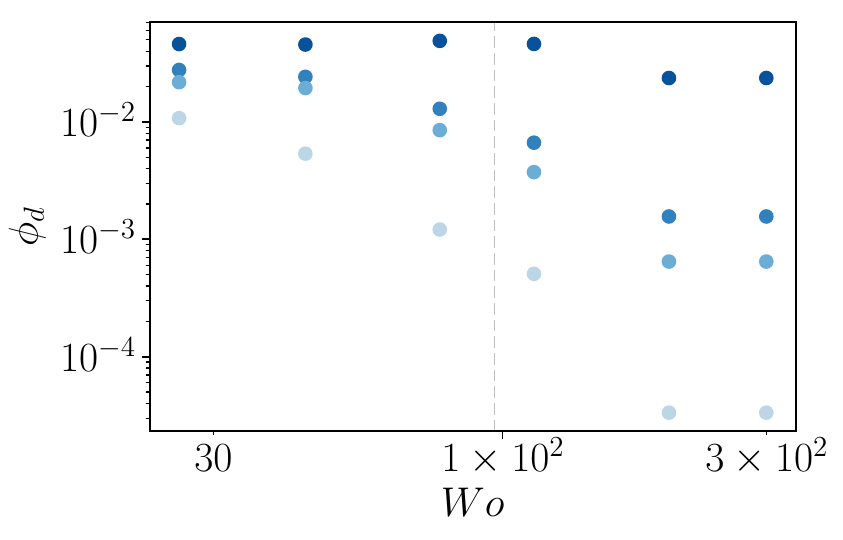}
\caption{Left panel: Amplitude response ($A_{u}$) against $Wo$ for non-rotating cases. The dashed line shows the scaling $A_u \sim Wo^{-2}$, and the inset shows the compensated amplitude $A_uWo^2$ against $Wo$ plot to emphasize the scaling. Right panel: Phase delay ($\phi_{d}$) using FFT against $Wo$ for non-rotating cases. Cases shown are $y=0.01, 0.05, 0.1. 0.5$ from light blue to dark blue.}
\label{fig:amp_vs_wo}
\end{figure}

To allow a more direct comparison to the results from \cite{VerschoofRubenA2018PdTt}, we plot the amplitude and phase lag as a function of $Wo$ for different $y$ locations in Figure \ref{fig:amp_vs_wo}. We have also indicated on both figures the line at which $T_\tau=1$. Unlike the corresponding figures in \cite{VerschoofRubenA2018PdTt}, this plot shows that the amplitude and phase delay of the perturbation is a strong function of the distance to the wall. There are two main possible sources for this discrepancy. First, we show values of $y$ which are much closer to the wall, down to $y=0.01$, while \cite{VerschoofRubenA2018PdTt} use distances which correspond to the range $0.2<y<0.8$. Second, not only the Reynolds number is different, but also the rotation rate. Taylor-Couette with a pure inner cylinder rotation corresponds to a Rotation number of $R_\Omega=1-\eta$, with $\eta$ being the radius ratio $\eta = r_i/r_o$, where $r_i$ is the inner cylinder radius and $r_o$ is the outer cylinder radius, respectively. This means that the effective value of $R_\Omega$ in the experiments is $R_\Omega=0.29$, as the radius ratio is $\eta=0.714$.
We will visit that value of $R_\Omega$ in a later section, and show that the effective solid-body rotation is indeed the main source of our discrepancy.

To summarize, there are two distinct regimes for $A_u(Wo)$: at low $Wo$, the amplitude is not a strong function of $Wo$. This region is especially pronounced for the data at $y=0.01$ for $Wo<100$. At high $Wo$, the amplitude rapidly decays as $Wo$ is increased. This decrease matches the prediction in \cite{von2003response67}: for high $Wo$, the amplitude should behave like $A_u\sim T$. This is shown in the figure as the dashed line $A_u\sim Wo^{-2}$ and in a compensated subplot. This theoretical scaling matches the data reasonably well. As we move away from the wall, the transition to the $A_u\sim Wo^{-2}$ dependence happens at lower values of $Wo$. We can attribute this to the flow finding it harder to adjust to the perturbation as the distance to the wall increases. This two region behaviour is also seen for the phase lag, shown in the right panel of Figure \ref{fig:amp_vs_wo}, even if no clear power law behaviour can be discerned, nor is it available from the theoretical derivations in \cite{von2003response67}.

%Fig 6
%% Au and Phase vs. Wo for y = 0.01, 0.05, 0.1, 0.5

\begin{figure}
\includegraphics[width=.45\linewidth]{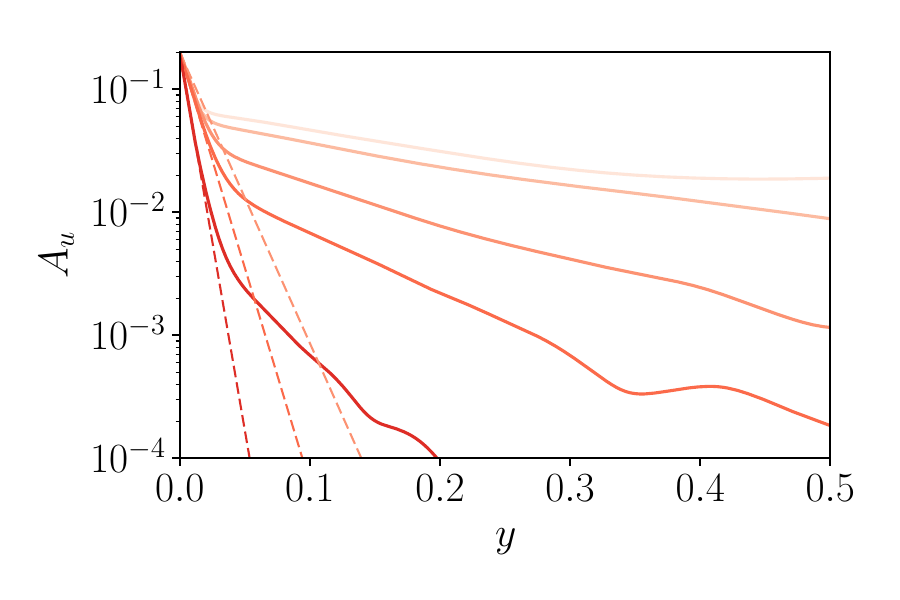}
\includegraphics[width=.45\linewidth]{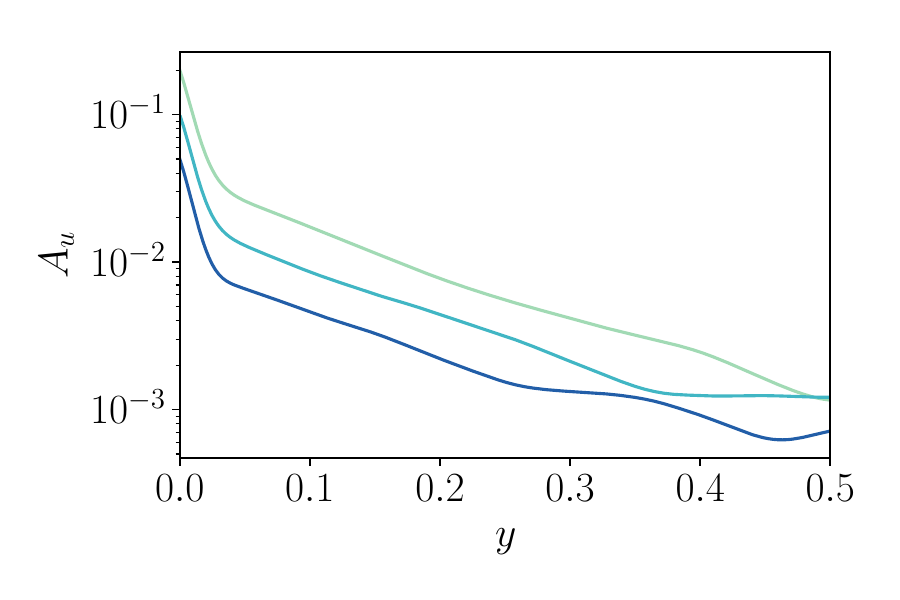}
\includegraphics[width=.45\linewidth]{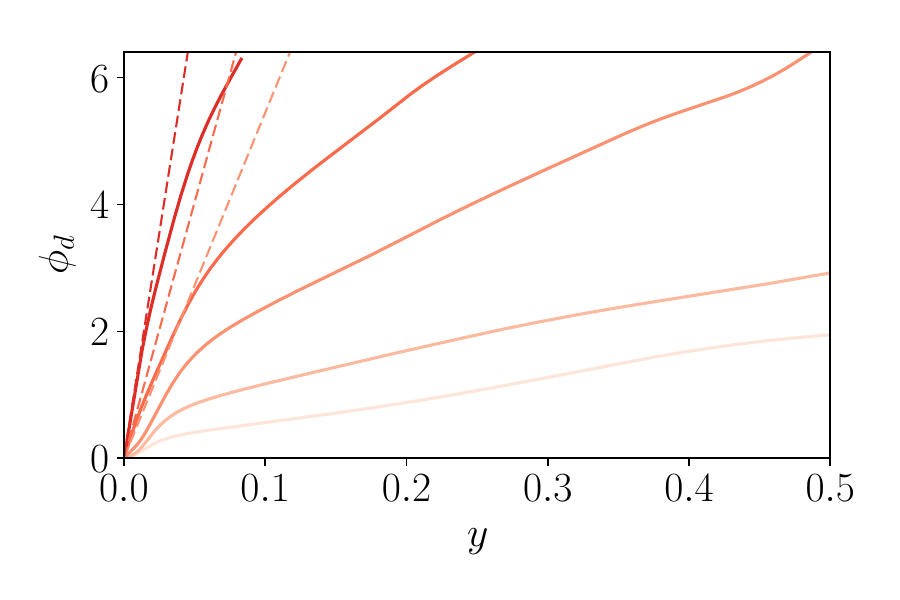}
\includegraphics[width=.45\linewidth]{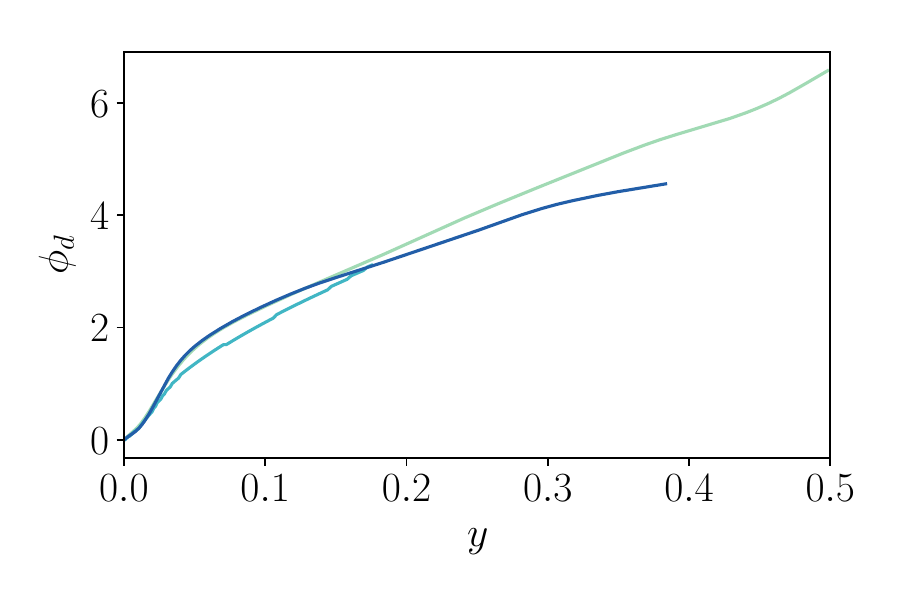}
\caption{Top-left panel: Amplitude response ($A_{u}$) against wall distance for $\alpha=0.2$ and varying $Wo$. Symbols are the same as Fig.~\ref{fig:phaselag}. Top right panel: Amplitude response against wall distance for $Wo=77$ and $\alpha=0.05$ (dark) , $\alpha=0.1$ (mid) and $\alpha=0.2$ (light). Bottom panels: same as top panels for phase delay ($\phi_d$).}
\label{fig:alphaeffect}
\end{figure}

For completeness, we checked the effect of the amplitude $\alpha$ on the results by running all the cases shown above for $\alpha=0.05$ and $\alpha=0.2$. A short summary of the results is presented in Figure \ref{fig:alphaeffect}. In the left panels we show the amplitude response and the phase delay against wall distance for all values of $Wo$ and $\alpha=0.2$. We can see the same qualitative phenomena we saw appear for $\alpha=0.1$, which we have already discussed. We do not show these results for $\alpha=0.05$ as they show the same patterns, but the numerical averaging errors appear for smaller values of $y$ due to the smaller amplitude of the perturbation. In the right panels of the figure, we show $A_u$ and $\phi_d$ against $y$ for different values of $\alpha$ and the same $Wo$. We can clearly see that the amplitude response of the system is simply offset by a factor, while the phase response is approximately independent of $\alpha$ except for some small discrepancies which we attribute to insufficient statistics.

\begin{figure}
\includegraphics[width=.45\linewidth]{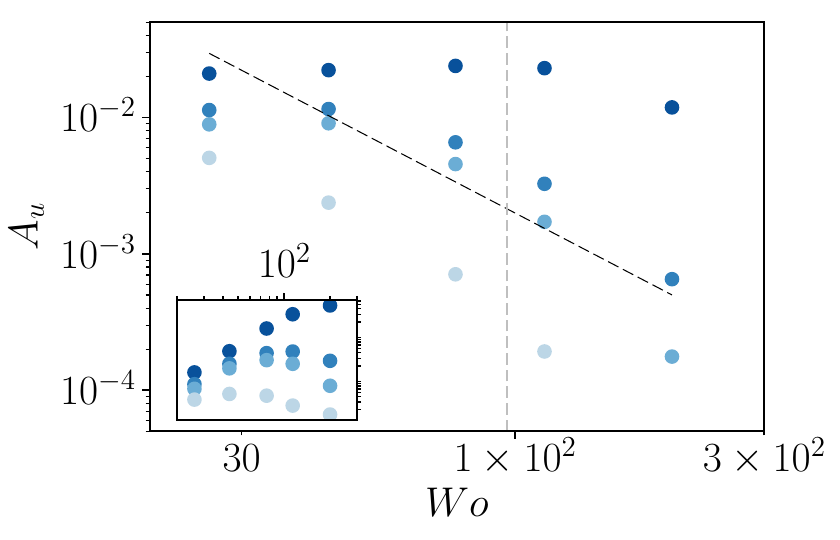}
\includegraphics[width=.45\linewidth]{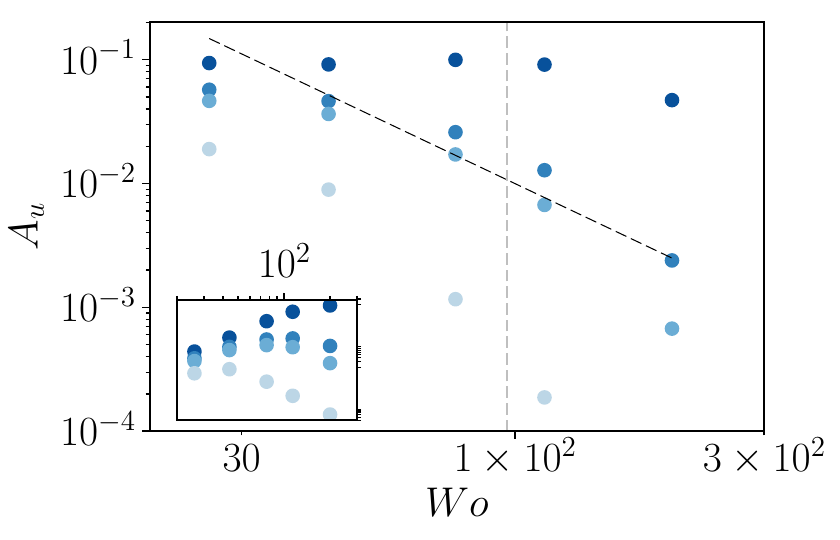}
\caption{Amplitude response ($A_{u}$) against $Wo$ for non-rotating cases and $\alpha=0.05$ (left) and $\alpha=0.2$ (right). The dashed line shows the scaling $A_u \sim Wo^{-2}$, and the inset shows the compensated amplitude $A_uWo^2$ against $Wo$ plot to emphasize the scaling. }
\label{fig:amp_vs_wo_alpha0p2}
\end{figure}

Finally, in Figure \ref{fig:amp_vs_wo_alpha0p2}, we show an analog to Figure \ref{fig:amp_vs_wo} but for the two other values of $\alpha$ simulated (0.05 and 0.2). It shows the same $A_u\sim Wo^{-2}$ behaviour in the high $Wo$ regime. This gives us confidence in the fact that for small values $\alpha$  is a physically unimportant parameter that is only relevant when considering the effect of numerical averaging errors. We can expect that for $\alpha\sim\mathcal{O}(1)$, significant effects of the amplitude modulation will begin to be seen in $Nu$, similar to those in \cite{yang2020periodically}.

\section{Results for rotating Plane Couette flow}

\subsection{Modulation and Taylor rolls}

Adding solid body rotation causes a drastic change in the flow behavior and modifies the underlying statistics such as dissipation and mean velocity \citep{brauckmann2016momentum}. As mentioned earlier, it also triggers the formation of large-scale pinned structures known as Taylor rolls which are primarily responsible for the transport of shear. These rolls are in close analog with the large-scale structures in Rayleigh-B\'enard flow that dominate heat transfer. Further, such large-scale structure couple to modulation introduced through the driving boundary (oscillating wall), \citep{jin2008experimental,yang2020periodically}.

\begin{figure}
 \includegraphics[width=.32\linewidth]{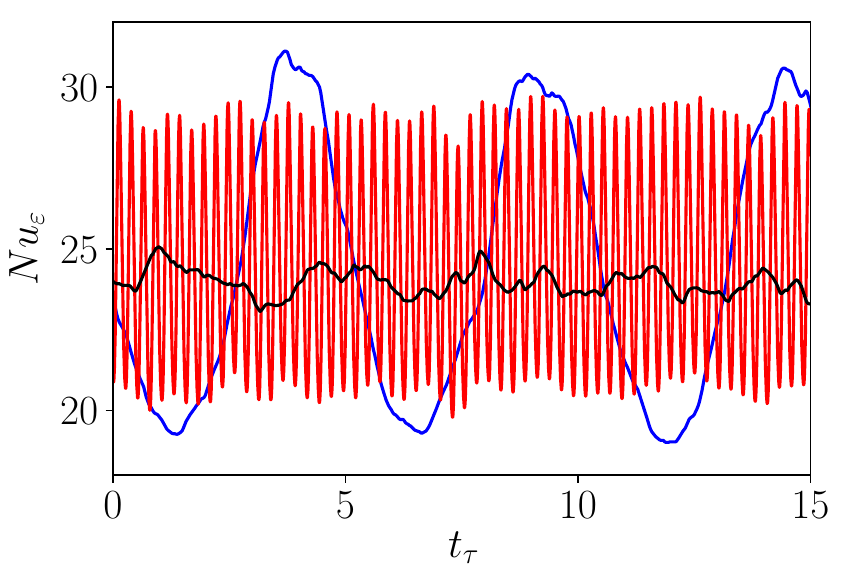}
 \includegraphics[width=.32\linewidth]{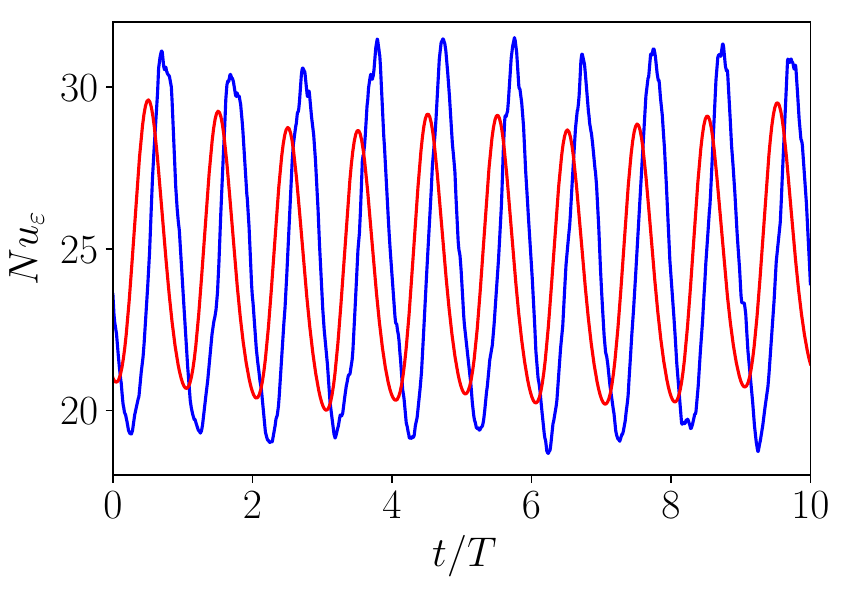}
 \includegraphics[width=.32\linewidth]{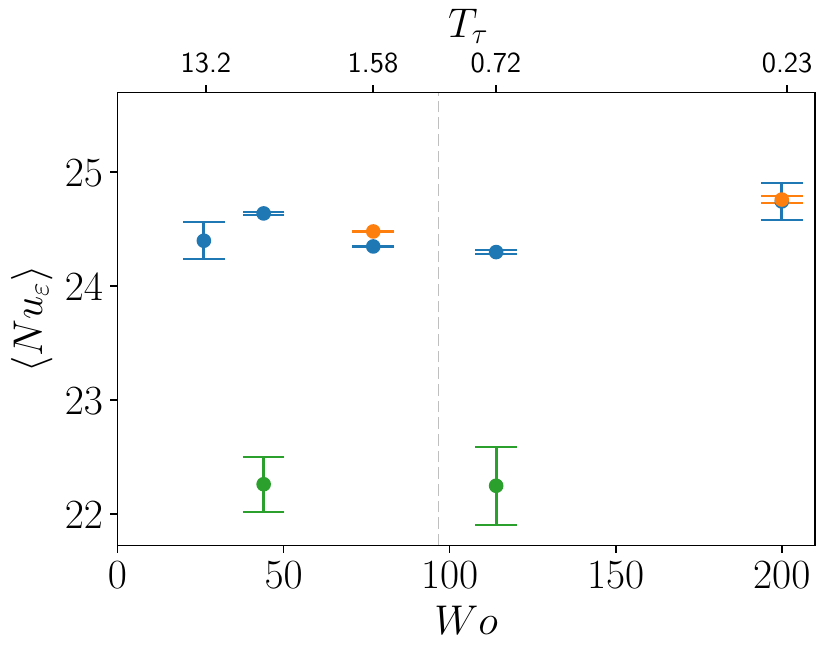}\\
 \caption{Left panel: Temporal evolution of the averaged dissipation, non-dimensionalized as a Nusselt number ($Nu$) for unmodulated flow (black), $Wo=44$ (blue) and $Wo=200$ (red) with $R_\Omega=0.1$. Center panel: Same as the left panel, with the time units re-scaled using the period of the forcing. Right panel: Temporally averaged $Nu$ against $Wo$ for $R_\Omega=0.1$. Blue points denote the baseline periodic aspect ratios of $L_x=2\pi$ and $L_z=\pi$ with a single roll pairs, while green points denote the same domain for two roll pairs. The orange data points are simulations to check the effect of domain size with $L_x=4\pi$ and $L_z=2\pi$ and two roll pairs (with the same roll size as the baseline).}
\label{fig:diss_ro01}
\end{figure}

We start our discussion with $R_\Omega=0.1$ which is the value of $R_\Omega$ for which the rolls are most energetic \citep{sacco2019dynamics}. Therefore, we can expect this to be the most favourable case to observe resonant coupling between the modulation and the existing structures in the flow, as the structures have very well defined natural length- and time-scales. We first show the instantaneous $Nu_\varepsilon(t)$ in the left and center panels of Figure \ref{fig:diss_ro01} for $R_\Omega=0.1$ and $Wo=44$, $Wo=200$ and unmodulated flow (black, only left panel). The temporal fluctuations due to the inherent turbulence of the flow are even smaller than for the case with no rotation. The modulation introduces fluctuations in $Nu_\varepsilon(t)$ which are of the order of $20-30\%$ of the mean value of $Nu_\varepsilon$, in the same order of magnitude as the values observed previously. We can also observe that the fluctuations in $Nu_\varepsilon$ are larger for smaller values of $Wo$, similar to what was observed in \cite{jin2008experimental}. We propose an explanation for this below. 

Similar to the non-rotating case, the mean value around which all curves fluctuate appears to be the same. To further quantify this, in the right panel of Figure \ref{fig:diss_ro01} we show the time-averaged values of $Nu_\varepsilon$ as a function of $Wo$. As was seen for the non-rotating case, no strong dependence of $Nu_\varepsilon$ with $Wo$ is observed. This hints at the fact that no significant resonances happen between the modulated forcing and the existing structure. These results are consistent with those obtained in \cite{jin2008experimental} for sinusoidally modulated RBC, who did not observe an enhancement in the time-averaged values of heat transport. We also confirm that the dependence of $Nu$ on domain size is much smaller for the rotating PCF.

%Fig 8
%% Nusselt vs t/T and Wo

We note that in all cases discussed here, we are analyzing simulations with a single roll pair, such that the roll wavelength $\lambda_{TR}=L_z=\pi$. The possibility for other roll states with different wavelengths to arise and persist in rotating PCF and TCF is well documented \citep{ostilla2016effect,xia2018multiple}. The number of rolls in a simulation is very dependent on the initial conditions, and through our choices we obtain simulations to have a single roll pair state. The adding of modulation at $\alpha=0.1$ is not strong enough to change the roll state in our selected domain size, and we did not see any effect of the modulation on these rolls. There still was a single roll pair in the averaged fields after turning the modulation on. By manipulating the initial conditions, we were able to generate states with two pairs of rolls for $Wo=44$ and $Wo=114$, such that the roll wavelength was now $\lambda_{TR}=L_x/2=\pi/2$. These show the same properties in the Nusselt number discussed above: larger oscillations with small $Wo$, and small oscillations with large $Wo$. However, the average Nusselt number value was lower, consistent with what was discussed in \cite{ostilla2014exploring}, where it was observed that for Taylor-Couette at similar values of $Re_s$, small rolls would produce lower values of the torque.

\begin{figure}
  \includegraphics[width=.48\textwidth]{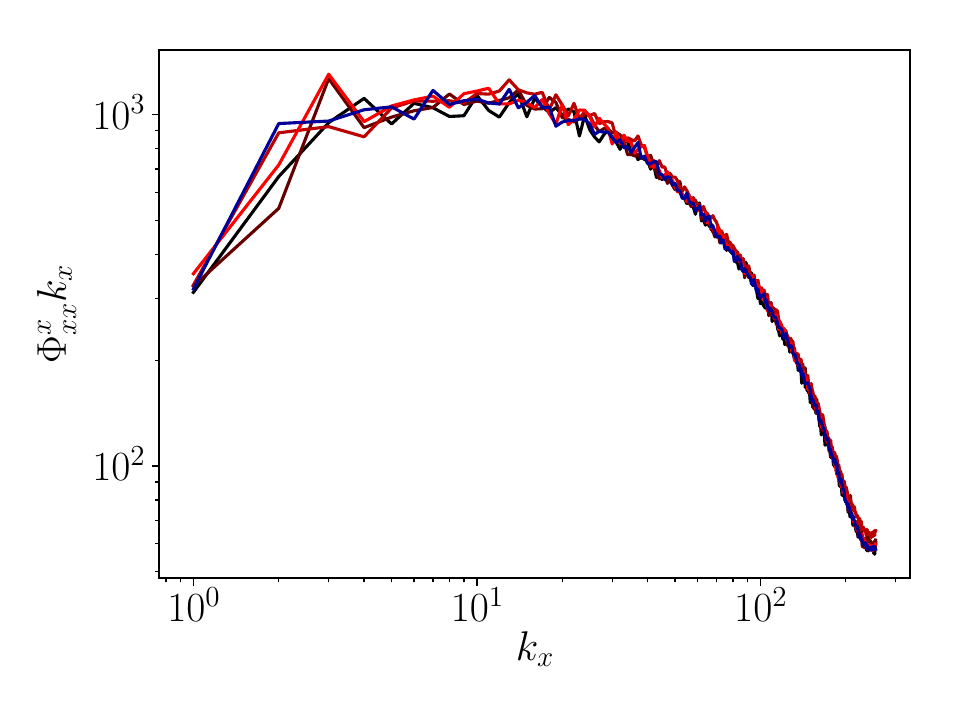}
  \includegraphics[width=.48\textwidth]{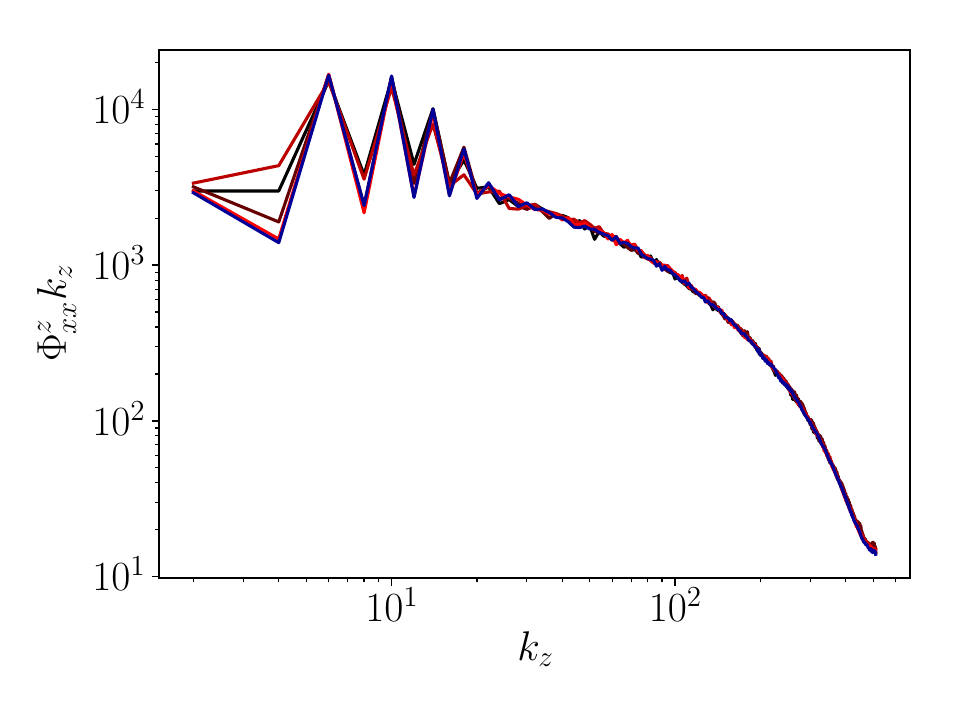}
\caption{Time-averaged pre-multiplied streamwise (left) and spanwise (right) spectra of the streamwise velocity at the mid-gap ($y=0.5$) for RPCF with $R_\Omega=0.1$. Symbols: $Wo=44$ (black) $Wo=77$ (dark red), $Wo=200$ (light red), unmodulated (dark blue).}
\label{fig:specro01}
\end{figure}

Further proof of the robustness of the rolls to modulation is provided by the velocity spectra. We show the energy spectrum of the streamwise velocity in Figure \ref{fig:specro01}. The ``sawtooth'' pattern characteristic of Taylor rolls (c.f.~\cite{ostilla2016near}) can be appreciated for the spanwise spectra $\Phi_{xx}^z$. Some modification of the pattern is observed for $Wo=77$, which is the value of $Wo$ that more closely matches the natural time-scales of the flow and is the one which we expect to produce resonant effects. However, this modification does not correspond to a significant change in the value of $\langle Nu_\varepsilon \rangle$. Instead we postulate that the modes which are normally dampened or eliminated by the Taylor rolls, i.e.~the second, fourth and other even fundamentals, are more energetic than for the unmodulated values. However, because the first fundamental mode remains unaffected, which corresponds to the Taylor roll and hence transports the most shear, the resulting transport of shear is almost unaffected. 

%Fig 9
%% Nusselt vs t/T and Wo

\begin{figure}
  \includegraphics[width=.48\textwidth]{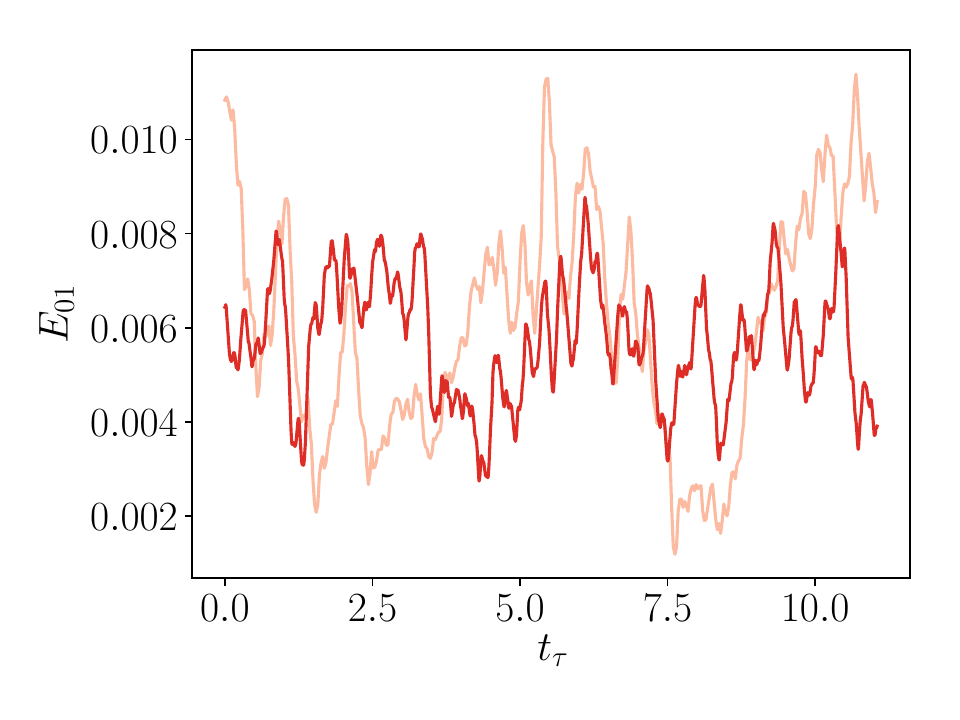}
    \includegraphics[width=.48\textwidth]{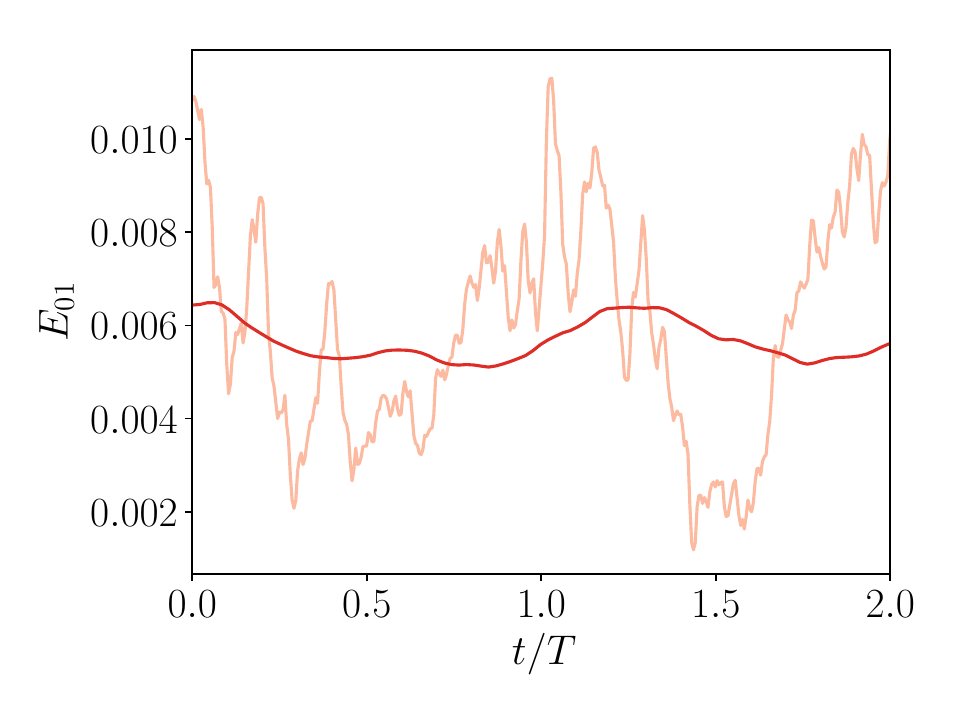}
\caption{Temporal evolution of the energy of the fundamental Fourier mode associated to the Taylor roll in flow (left) and forcing (right) time units for $R_\Omega=0.1$. Symbols: light red lines $Wo=44$, dark red lines $Wo=200$.}
\label{fig:evst}
\end{figure}

Following the discussion in \cite{jin2008experimental}, it could be that sinusoidal modulations are too smooth to cause a resonance sufficient to increase the strength of the roll and hence change $\langle Nu_\varepsilon \rangle$. We can justify this by looking at the energy of the Taylor roll. Following \cite{sacco2019dynamics}, we measure this energy as the energy $E_{01}$ of the first $z$-fundamental (i.e.~$k_x=0$, $k_z=2$) of the wall-normal velocity at the mid-gap. In Figure \ref{fig:evst} we show the temporal behaviour of this quantity. For $Wo=200$, the inertia of the roll is large enough to absorb the forcing modulation. Conversely, for $Wo=44$, the time-scale of the modulation becomes larger than $t_\tau$. The roll's inertia is insufficient to absorb this slow modulation, and its signature appears as a larger fluctuations in $E_{01}$. This explains the larger flucutations observed in $Nu(t)$ for small values of $Wo$ at $R_\Omega$.

Moreover, the lack of large-scale shear-transporting structures at $R_\Omega=0$ means we do not see an increased fluctuation level in Fig.~\ref{fig:diss}  We can further justify this by looking at $E_{01}$ for the non-rotating case. We show the time evolution of this quantity in Fig.~\ref{fig:evst}. We can clearly see that there is no obvious correlation between forcing and $E_{01}$, and the energy of the mode changes in time with no discernable pattern. This is further proof that while the modulated drivings can affect large scale structures, sinusoidal drivings cannot achieve transport-enhancing resonances.

\begin{figure}
  \includegraphics[width=.48\textwidth]{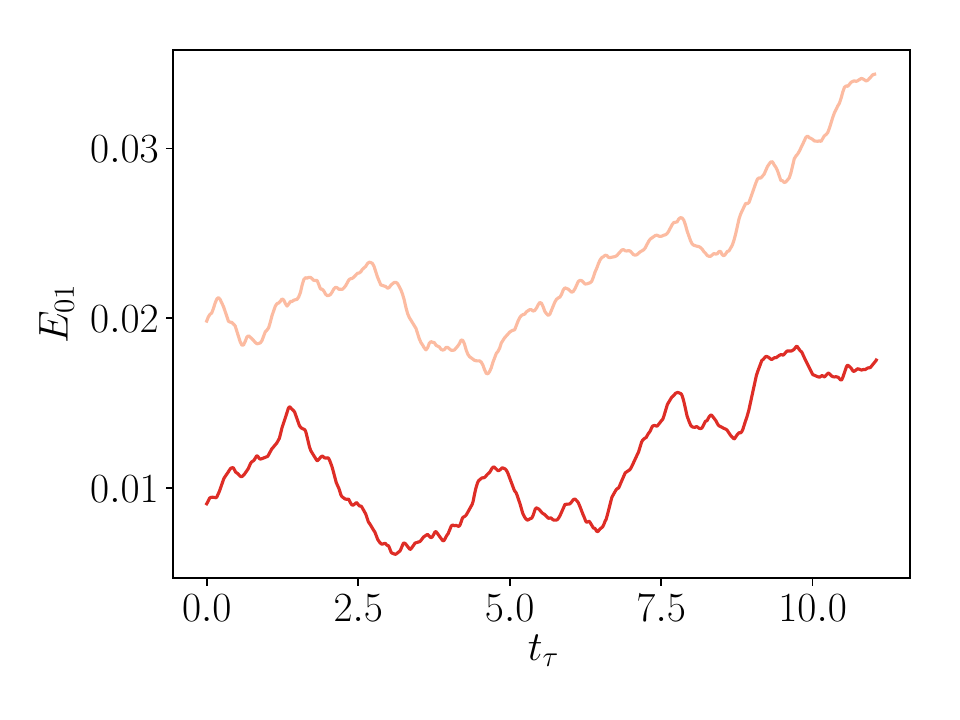}
    \includegraphics[width=.48\textwidth]{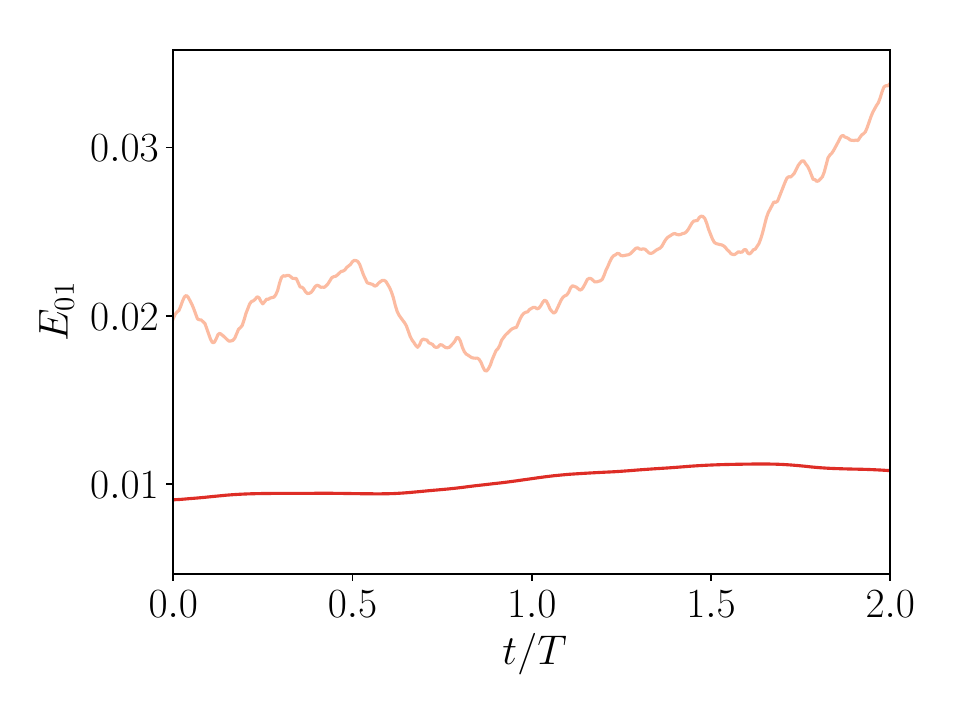}
\caption{Temporal evolution of the energy of the fundamental Fourier mode associated to the Taylor roll in flow (left) and forcing (right) time units for non-rotating PCF. Symbols: light red lines $Wo=44$, dark red lines $Wo=200$.}
\label{fig:evstRo0}
\end{figure}

Unfortunately, the presence of the roll prevents us from examining the way the modulated energy propagates into the flow, as the introduced spanwise dependence of average quantities makes replicating the earlier analysis of $\S$\ref{sec:dissro0} impossible. The decomposition introduced to calculate $f$ as in Eq.~\ref{eq:fdefin} produces large artifacts which we will discuss later, in context with the results from other values of $R_\Omega$.

\subsection{Modulation and mean rotation }

\begin{figure}
 \includegraphics[width=.49\linewidth]{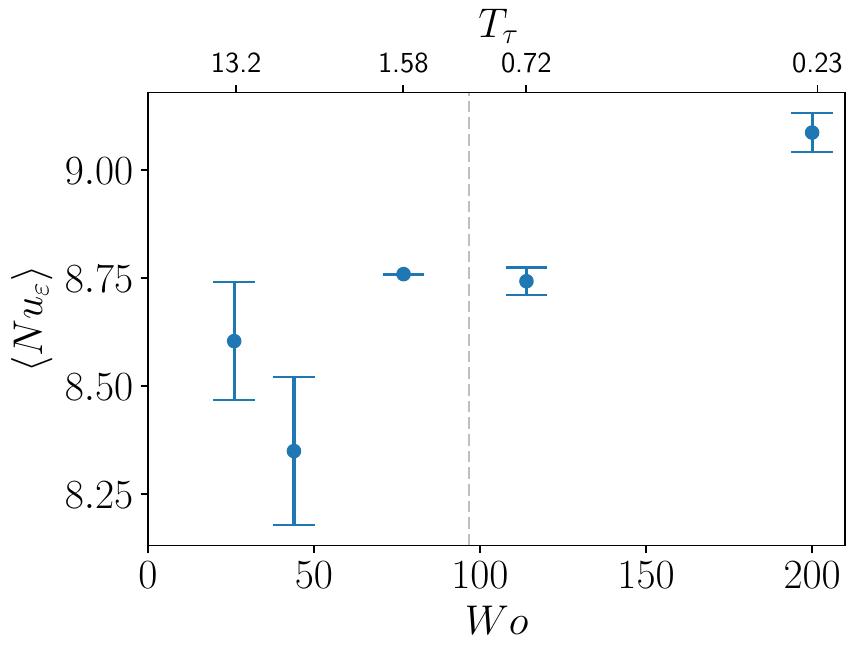}
 \includegraphics[width=.49\linewidth]{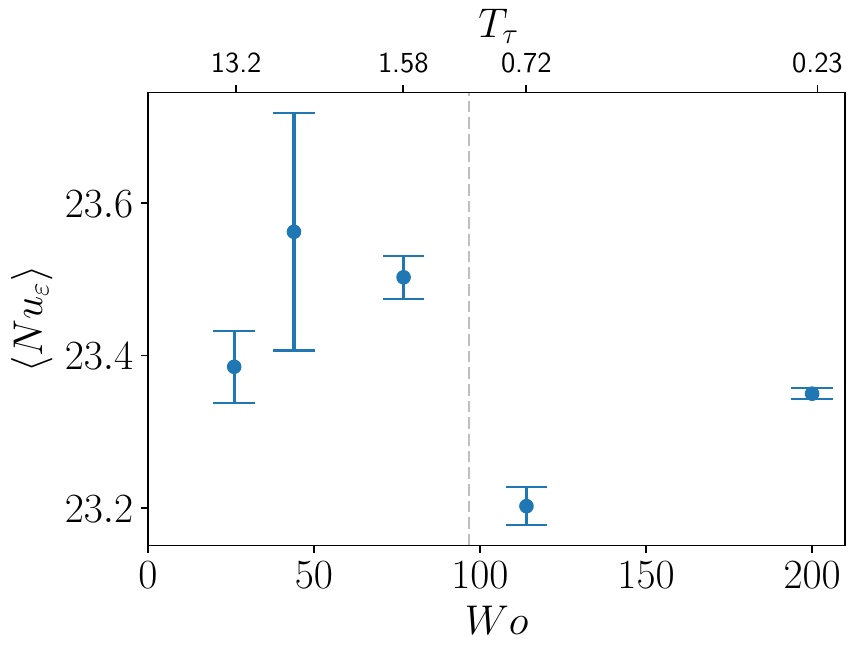}
 \caption{Temporally averaged $\langle Nu_\varepsilon \rangle$ against $Wo$ for $R_\Omega=-0.1$ (left) and $0.3$ (right).}
\label{fig:diss_ros}
\end{figure}

We start by analyzing the effect the remaining values of $R_\Omega$ have on the behaviour of the shear transport and the dissipation. 
In Figure \ref{fig:diss_ros} we show the time-averaged $Nu_\varepsilon$ against $Wo$ for the two remaining values of $R_\Omega$ studied. As for the previous cases, no significant effect of $Wo$ can be appreciated on the average value of $Nu_\varepsilon$. We may now conclude that there is no significant dependence of $Nu_\varepsilon$ on $Wo$ across all our simulations, and that the modulation does not significantly modify the average shear transport.

\begin{figure}
\includegraphics[width=.49\linewidth]{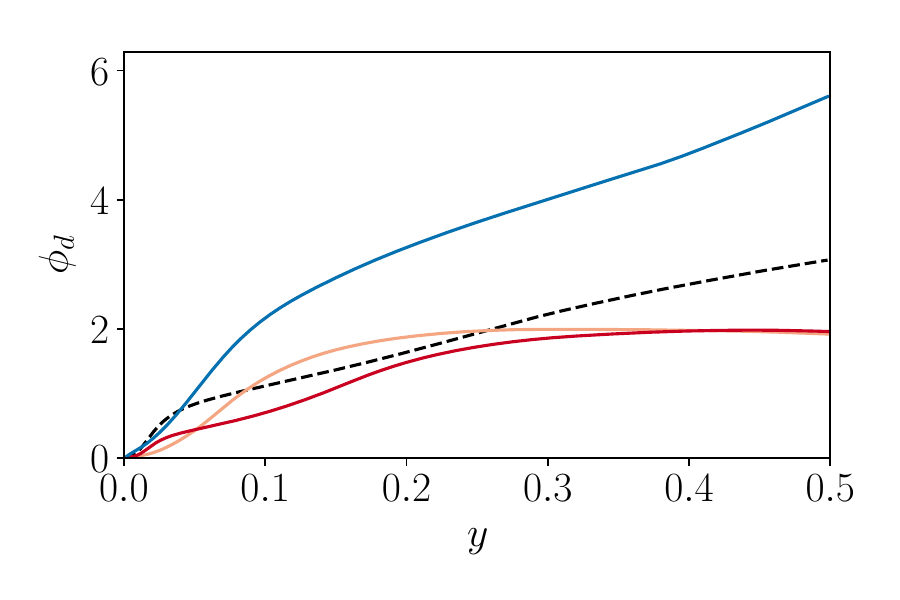}
\includegraphics[width=.49\linewidth]{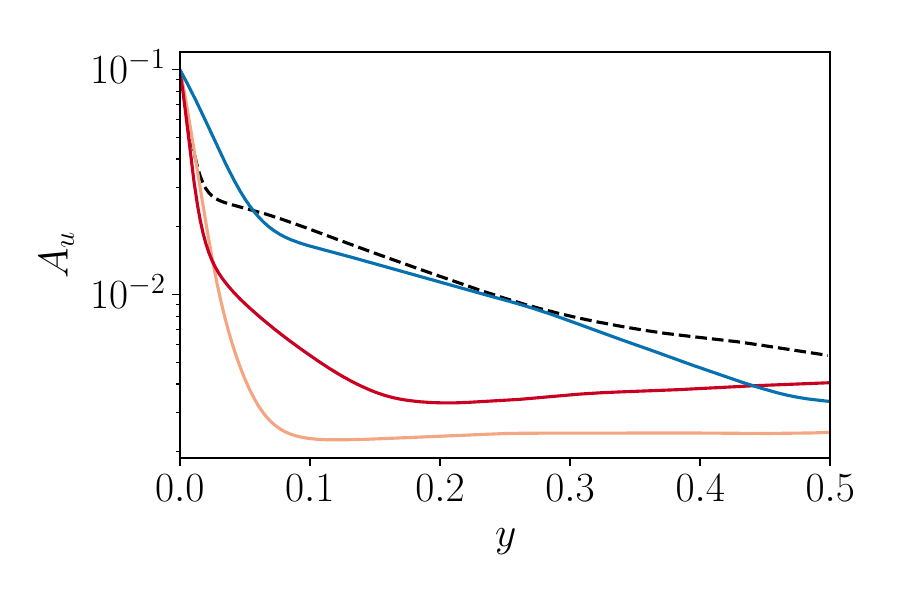}
\caption{Left: Phase delay ($\phi_{d}$) against wall distance. Right: Amplitude response ($A_u$) against wall distance. Symbols: $R_\Omega = 0$ (dashed black), $0.1$ (light red), $0.3$ (dark red), $-0.1$ (dark blue). $Wo=44$ for all cases. }
\label{fig:response_ros}
\end{figure}

We now turn to the way the modulation propagates into the flow. We use the decomposition from Eq.~\ref{eq:fdefin} to calculate the amplitude response and phase delay against wall distance. We show the results obtained, alongside those for the non-rotating case, in Figure \ref{fig:response_ros}. Significant differences between all curves can be seen. The case with cyclonic rotation ($R_\Omega=-0.1$) follows a similar pattern as the case with no rotation. Close to the wall in the viscous sub-layer, the modulation is transported at a slow and constant velocity (reflected as a linear behaviour of the phase delay), with an exponential decay of the amplitude response. As the distance is increased, a second region appears where the perturbation also travels at a constant but larger velocity, and decays exponentially but with a smaller exponent, equivalent to the flow gaining a larger effective viscosity. The transition between these regions is located at $y\approx0.08$, slightly larger than the transition between the regions seen for the non-rotating case. We can attribute this to the lower levels of turbulence and to the lower frictional velocities present when cyclonic rotation is added, which means that the viscous sub-region extends further from the wall (c.f.~\cite{ostilla2016turbulent} for an analysis of Taylor-Couette with only outer cylinder rotation, which has an equivalent cyclonic solid body rotation of $R_\Omega=-0.1$). Indeed, for $R_\Omega=-0.1$, $y=0.08$ corresponds to $y^+\approx 40$, in the buffer region.

In contrast, the results for anti-cyclonic rotation ($R_\Omega>0$) do present more differences when compared to the non-rotating case. For $R_\Omega=0.1$, the amplitude response drops very rapidly as we move away from the wall, and then remains constant from $y\ge 0.08$, while the phase delay also increases at first, and then remains constant from $y \ge 0.15$. The transition between one type of behaviour to the other happens at very different values of $y$, and the behaviour is very different from the constant velocity or constant exponential decay seen for $R_\Omega \leq 0$, and also deviates strongly from the behaviour of Stokes' solution, namely linear phase increases and exponential amplitude decay. As we will further justify below, we can attribute this strange behaviour to the spanwise inhomogeneities introduced by the Taylor rolls, which introduce numerical artifacts and affect the calculation of $f$.

Turning to $R_\Omega=0.3$, we can observe the two near-wall and bulk regions which have a linear behaviour for the phase delay and an exponential decaying behaviour for the amplitude response, as well as a third region for $y\ge 0.2$ which was previously unobserved and where both the phase delay and the amplitude response are constant. This indicates that rotation induces a region that is affected simultaneously and homogeneously by the flow modulations. Unlike for $R_\Omega=0.1$, we believe that these results are not due to averaging artifacts, as they were also observed in the experiments in \cite{VerschoofRubenA2018PdTt}, who observed no significant dependence of $A_u$ and $\phi_d$ on $y$. As mentioned earlier, the setup in \cite{VerschoofRubenA2018PdTt} corresponds to an effective $R_\Omega=0.29$, very close to our simulated value, and furthermore, the experiment only measured $A_u$ and $\phi_d$ at distances from the wall of $0.2<y<0.8$. According to our simulations, these distances would all correspond to the region where $A_u$ and $\phi_d$ are constant. This explains the discrepancy between our earlier results and the experiment: anti-cyclonic rotation adds a new physical phenomena where the perturbation coming from the wall modulation appears to be constant throughout the bulk. 

%% Au and Phase vs. y for Ro = 0, 0.1, 0.3, -0.1

\begin{figure}
\includegraphics[width=.32\linewidth]{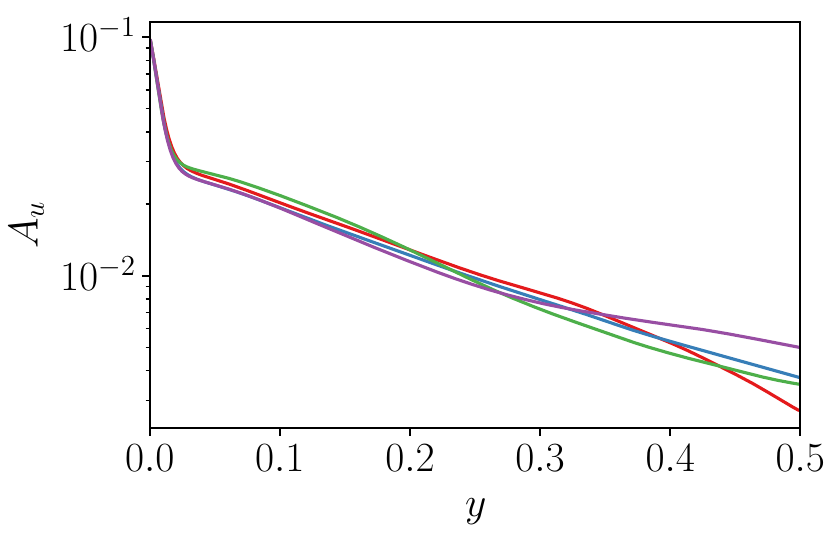}
\includegraphics[width=.32\linewidth]{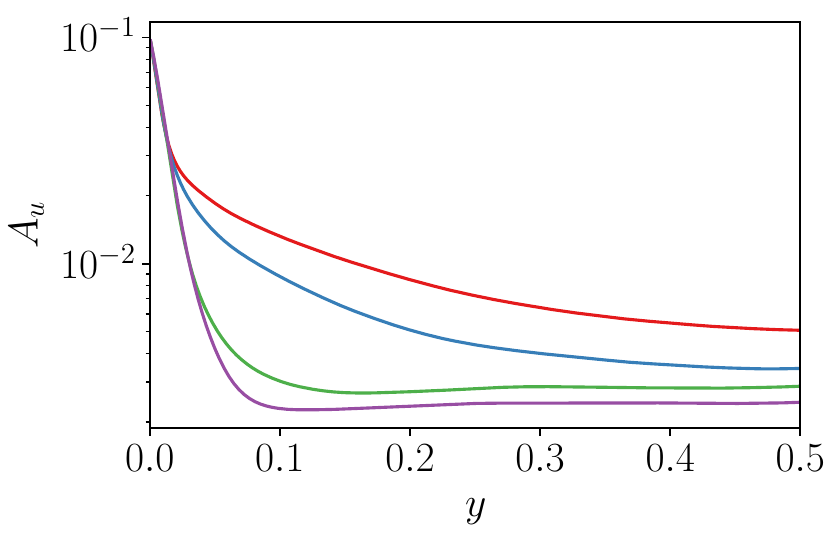}
\includegraphics[width=.32\linewidth]{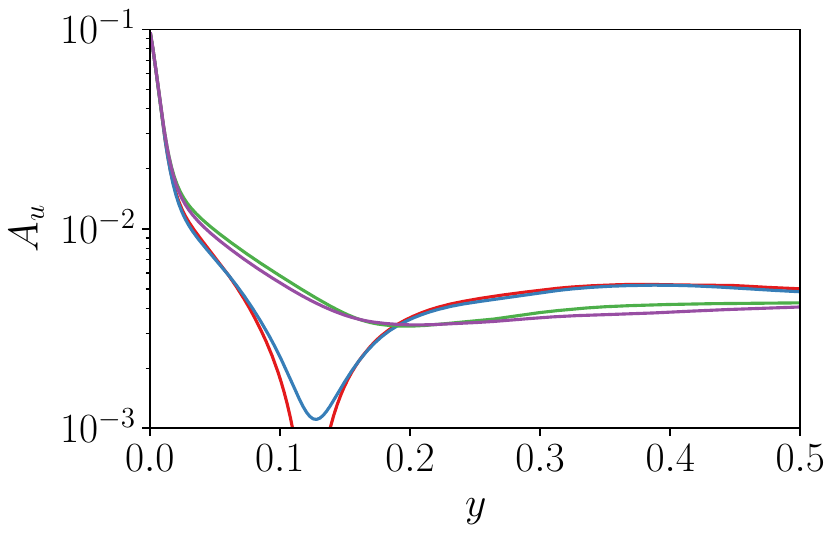}
\includegraphics[width=.32\linewidth]{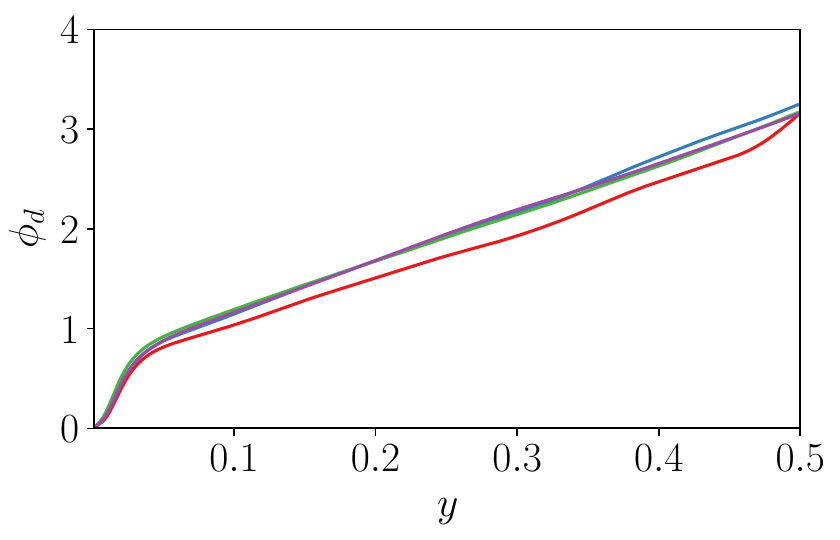}
\includegraphics[width=.32\linewidth]{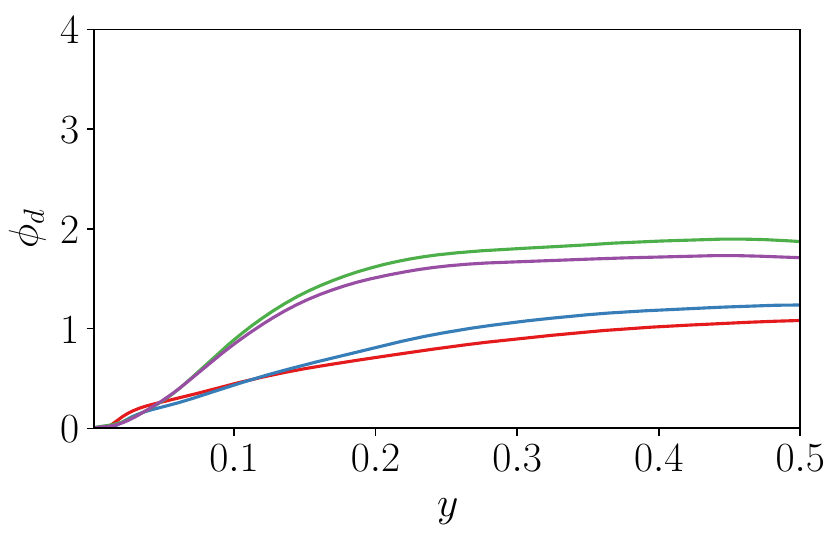}
\includegraphics[width=.32\linewidth]{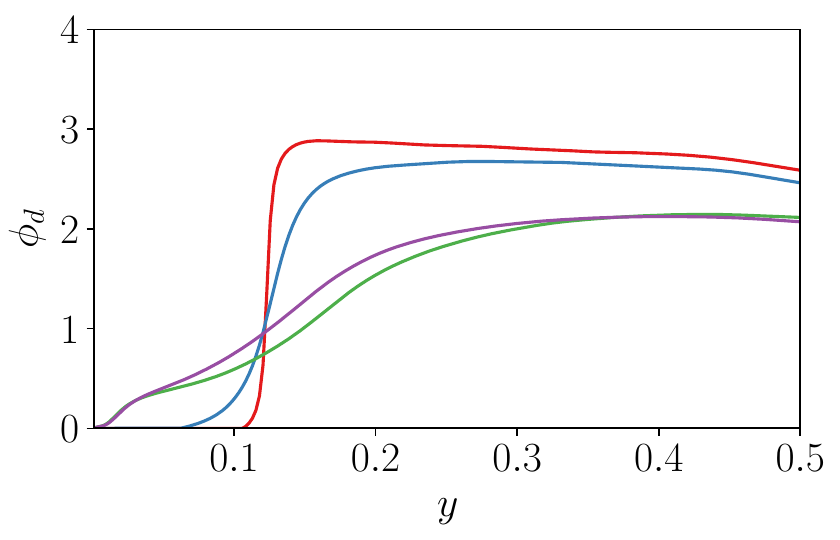}
\caption{Amplitude response ($A_{u}$, top row) and Phase delay ($\phi_d$, bottom row) against wall distance for $Wo=44$ and $R_\Omega=0$ (left column), $R_\Omega=0.1$ (center column) and $R_\Omega=0.3$ (right column). The different lines denote different averaging windows for azimuthal velocity: $L_z/8$  (red), $L_z/4$ (dark blue), $L_z/2$ (green) and full $L_z$ extent (purple)}
\label{fig:window}
\end{figure}

To further justify that the presence of large-scale structures interferes with the measurement of the perturbation amplitude and phase delay for $R_\Omega=0.1$, in figure \ref{fig:window} we show the effects of calculating $A_u$ and $\phi_d$ by only averaging over a fraction of the span-wise length. The non-rotating case is used as a baseline which shows that even taking an eighth of the spanwise domain length as an averaging window leads to reasonable results if large-scale structures are not present. Once the rotation is increased to $R_\Omega=0.1$, it is impossible to obtain results which do not depend on the extent of the spanwise averaging window. Increasing the domain size leads to the amplitude response to drop rapidly for the same $y$ coordinate, as the effect of spanwise inhomogeneity contaminates the measurement of $f$. Finally, when $R_\Omega$ is further increased to $0.3$, the Taylor rolls disappear, and while neither an eighth of the domain nor a quarter is enough to produce accurate results, the results from half- and the full domain show very similar behaviour. We can thus use our results for $R_\Omega=0.3$ with confidence.

%% Window size dependence on Au and Phase vs. y for Ro = 0, 0.1, 0.3

\begin{figure}
\includegraphics[width=.45\linewidth]{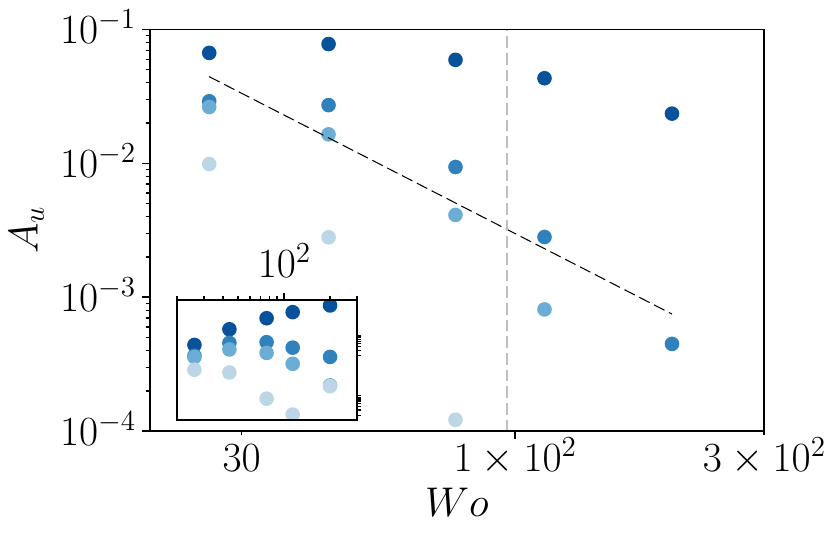}
\includegraphics[width=.45\linewidth]{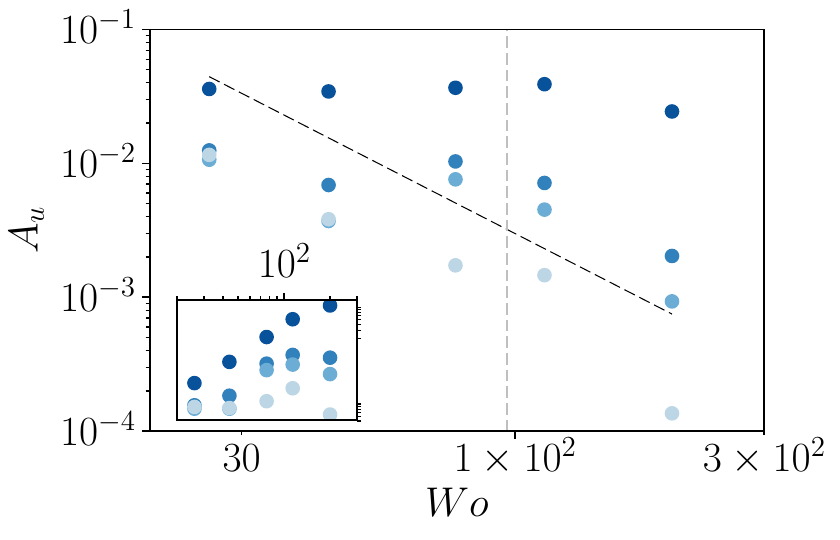}
\caption{Amplitude response ($A_{u}$) against $Wo$ for $\alpha=0.1$ and $R_\Omega=-0.1$ (left) and $R_\Omega=0.3$ (right). The dashed line shows the scaling $A_u \sim Wo^{-2}$, and the inset shows the compensated amplitude $A_uWo^2$ against $Wo$ plot to emphasize the scaling. }
\label{fig:amp_vs_wo_ros}
\end{figure}

We finalize this section by showing in Figure \ref{fig:amp_vs_wo_ros} the amplitude response as a function of $Wo$ for several wall distances at $R_\Omega=-0.1$ and $0.3$. For $R_\Omega=-0.1$ we can observe very similar results to those seen for the non-rotating case, with a small region consistent with $A_u\sim Wo^{-2}$ behaviour. This shows that for cyclonic rotation, the physics of the flow's response to modulation remains largely unchanged.

For $R_\Omega=0.3$, we cannot observe any behaviour consistent with power-laws. We can compare this case to the experiments of \cite{VerschoofRubenA2018PdTt}, as we have some similarities. First, we can observe a similar collapse of $A_u$ for low $Wo$ at $0.05<y<0.5$, as well as a slow divergence of the curves as $Wo$ becomes larger. While in \cite{VerschoofRubenA2018PdTt} this collapse held until $Wo>100$, in our case the data can only really be seen to collapse for $Wo=26$. Furthermore, the experiments reported an exponential-like decay for $A_u$ as a function of $Wo$, but we do not observe behaviour consistent with exponential decay. The differences between the experiment and the simulations can probably be attributed to two sources: first, the experiments do not report data very close to the wall, so a fair comparison would not include the data points in the darkest blue. Second, the correct way to express the dimensionless period is not the viscosity-based $Wo$ but $T_\tau$, as the flow is fully turbulent. Therefore, it does not make sense to compare cases at the same $Wo$, but instead we should compare cases and flow behaviour transitions for the same values of $T_\tau$. In our simulations $Wo\approx 80$ corresponds to $T_\tau=1$, while in their experiments we can expect this number to be closer to $Wo\approx 220$. Therefore, the separation of data we see at $Wo\sim40$ approximately corresponds to the separation of curves observed at $Wo\sim110$ in the experiments, making the observation of $y$-independence for the amplitude response at large forcing periods consistent across simulations and experiments.

\section{Summary and Outlook}
% New methods to calculate the perturbation
% track it
We performed direct numerical simulation (DNS) of non-rotating and rotating Plane Couette flow with a modulated plate at a fixed shear Reynolds number, $Re_s= 3 \times 10^4$, for Womersley numbers in the range $Wo \in [26,200]$ while keeping the amplitude of the modulation constant at $\alpha=0.1$. We also studied the effect of cyclonic and anti-cyclonic Coriolis forces in the system by varying the rotation parameter, $R_\Omega$ in the range $\in [-0.1, 0.3]$. 

The average shear at the walls and the instantaneous dissipation was found to be independent of the modulation frequency regardless of the Coriolis force added, and no evidence of resonance between flow structures and modulation was found, consistent with the RBC results of \cite{jin2008experimental}.

The propagation of the modulation was measured using Fourier transforms of phase-averaged velocities to obtain the phase delay and amplitude response. For the non-rotating case ($R_\Omega = 0$), both the modulation response amplitude and the phase delay show behaviour similar to the theoretical solution of Stokes' boundary problem, i.e. a linear behaviour for the phase delay and an exponential decay for the amplitude response. There are two main regimes, a near-wall regime where the effective viscosity is close to the fluid viscosity, and a bulk regime where the effective viscosity appears to be much higher. The transition between slopes is observed at $y\approx0.05$, which corresponds to $y^+ = 40$ in viscous units, i.e. the transition between the viscid- and log-law regions of the turbulent boundary layer. When plotting the amplitude response as a function of $Wo$, we found a high-$Wo$ regime with behaviour consistent with the $A_u\sim Wo^{-2}$ behaviour as is expected of modulated turbulence at high frequencies \citep{von2003response67}. We also confirmed that the amplitude of the modulation is an unimportant parameter in determining the physics of the system for $\alpha<0.2$.

The simulations with cyclonic rotation result in similar behaviour to that of the non-rotating case: the modulation amplitude falls off exponentially at two different slopes with the change of slope now occurring at $y\approx0.08$, which again happens to be at $y^+ = 40$--the transition between the viscous sub-layer and the buffer layer. Results for anti-cyclonic conditions were presented for $R_\Omega = 0.1$ and $0.3$. For both anti-cyclonic conditions the amplitude decay as a function of distance from the wall exhibits very different behavior than the non-rotating case or cyclonic conditions and deviates from Stokes' solution. We attributed this marked difference at $R_\Omega=0.1$ to the presence of Taylor rolls, which introduce spanwise inhomogeneities and prevent the adequate calculation of the amplitude response. Furthermore, for $R_\Omega=0.3$, the amplitude decay and phase delay at $y\geq0.2$ remains fairly constant and is consistent with the observations in \cite{VerschoofRubenA2018PdTt}. We conclude by noting that the correct non-dimensional time-scale of modulated and rotating turbulent Couette flow is the frictional time-scale $t_\tau$, and expressing the period as $T_\tau$ leads to good correspondence of the amplitude response behavior between these current simulations and the experimental observations of \cite{VerschoofRubenA2018PdTt}.

The behaviour of the anti-cyclonic cases warrants further investigation, especially that seen at $R_\Omega=0.3$ which corresponds physically to a bulk region which feels the modulation all at once. Other types of modulation, such as periodic pulses should also be analyzed to find whether the average shear at the plates can be modified. To investigate these cases, more advanced ways of studying how the modulation propagates into the fluid must be developed, as the simple phase-averaged Fourier transform will not be able to produce adequate results.  

\noindent {\it Acknowledgments:} We acknowledge the Research Computing Data Core, RCDC, at the University of Houston for providing us with computational resources and technical support.

\noindent {\it Declaration of Interests:} The authors report no conflict of interest.

%Bibliographic references
%\bibliographystyle{abbr}
% \clearpage
\bibliographystyle{jfm}
% Note the spaces between the initials
%\bibliography{jfm-instructions}
\bibliography{apssamp}

\end{document}